\documentclass[%
 reprint,
superscriptaddress,
 amsmath,amssymb,
 aps,
prb,
]{revtex4-1}

\usepackage{bm}
\usepackage{graphicx}
\usepackage{hyperref}

\usepackage{xcolor}

\newcommand{\lam}{\lambda_\mathrm{ab}}
\newcommand{\Lam}{$\lam$}
\newcommand{\fp}{F_\mathrm{pin}}
\newcommand{\Fp}{$\fp$}
\newcommand{\flatmax}{F_\mathrm{lateral}^\mathrm{max}}
\newcommand{\Flatmax}{$\flatmax$}

\newcommand{\Lamz}{$\lam^{0\mathrm{K}}$}
\newcommand{\micro}[1]{$\mathrm{\mu}${#1}}
\newcommand{\um}{\micro{m}}

\newcommand{\BaFeAs}{$\mathrm{BaFe_2As_2}$}
\newcommand{\CoBa}{$\mathrm{Ba(Fe_{1-x}Co_x)_2As_2}$}
\newcommand{\PBa}{$\mathrm{BaFe_2(As_{1-x}P_x)_2}$}
\newcommand{\KBa}{$\mathrm{Ba_{1-x}K_xFe_2As_2}$}

\newcommand{\YBCO}{$\mathrm{YBa_2Cu_3O_{7-\delta}}$}

\newcommand{\tc}{T_C}
\newcommand{\Tc}{$\tc$}

\newcommand{\fig}{Fig.~}
\newcommand{\figs}{Figs.~}
\newcommand{\Fig}[1]{\fig\ref{#1}}
\newcommand{\Figs}[1]{\figs\ref{#1}}

\newcommand{\etal}{\textit{et al.}}

\newcommand{\xopt}{x_\mathrm{opt}}
\newcommand{\Xopt}{$\xopt$}

\begin{document}

\title{The dependence of the absolute value penetration depth on doping in $\mathrm{\mathbf{{(Ba_{1-x}K_x)Fe_2As_2}}}$}

\author{Avior Almoalem}
\affiliation{Department of Physics, Technion -- Israel Institute of Technology, Haifa, 32000, Israel}
\author{Alon Yagil}
\affiliation{Department of Physics, Technion -- Israel Institute of Technology, Haifa, 32000, Israel}
\author{Kyuil Cho}
\affiliation{Ames Laboratory, Ames, IA 50011, USA}
\affiliation{Department of Physics and Astronomy, Iowa State University, Ames, IA 50011, USA}
\author{Serafim Teknowijoyo}
\affiliation{Ames Laboratory, Ames, IA 50011, USA}
\affiliation{Department of Physics and Astronomy, Iowa State University, Ames, IA 50011, USA}
\author{Makariy A. Tanatar}
\affiliation{Ames Laboratory, Ames, IA 50011, USA}
\affiliation{Department of Physics and Astronomy, Iowa State University, Ames, IA 50011, USA}
\author{Ruslan Prozorov}
\affiliation{Ames Laboratory, Ames, IA 50011, USA}
\affiliation{Department of Physics and Astronomy, Iowa State University, Ames, IA 50011, USA}
\author{Yong Liu}
\affiliation{Ames Laboratory, Ames, IA 50011, USA}
\author{Thomas A. Lograsso}
\affiliation{Ames Laboratory, Ames, IA 50011, USA}
\affiliation{Department of Materials Science and Engineering, Iowa State University, Ames, IA 50011, USA}
\author{Ophir M. Auslaender}\email[]{ophir@physics.technion.ac.il}
\affiliation{Department of Physics, Technion -- Israel Institute of Technology, Haifa, 32000, Israel}

\begin{abstract}
We report magnetic force microscopy (MFM) measurements on the iron-based superconductor \KBa. By measuring locally the Meissner repulsion with the magnetic MFM tip, we determine the absolute value of the in-plane magnetic penetration depth (\Lam) in underdoped, optimally-doped, and overdoped samples. The results suggest an abrupt increase of \Lam\ as doping is increased from \Xopt, which is potentially related to the presence of a quantum critical point. The response of superconducting vortices to magnetic forces exerted by the MFM tip for $x=0.19$ and $0.58$ is compatible with previously observed structural symmetries at those doping levels.
\end{abstract}

\maketitle

\section{\label{sec:intro}Introduction}
Many aspects of superconductivity in the iron-based superconductors (FeSCs) are still not well understood. These materials exhibit novel phenomena such as the coexistence of magnetism and superconductivity \cite{Luan2011, Avci2011, Boehmer2015, Lamhot2015, Reid2016}, as well as more exotic behavior \cite{Hashimoto2012,Lamhot2015,Yagil2016}. One family with a particularly intriguing phase diagram is \BaFeAs, of which \KBa\ is a member. Here we report spatially resolved local measurements of the superconducting phase itself, and its relation with structural phases through vortex position and motion.

The phase diagram of \KBa\ shares many features with the phase diagrams of other pnictides [e.g. \CoBa \cite{Pratt2009, Nandi2010}, \PBa\  \cite{Kasahara2012, Bohemer2012, Iye2012a, Iye2012b,Hu2015}]. The parent compound, \BaFeAs, is a multiband metal that undergoes magnetic and structural phase transitions at $T_N \approx T_S\approx135$~K \cite{Avci2011, Boehmer2015}. Upon doping $T_{N,S}$ are suppressed until they vanish near $x\approx0.28$ \cite{Boehmer2015}. The system becomes superconducting at $T<T_C(x)$ for $x\gtrsim0.15$ \cite{Boehmer2015, Avci2011}. $T_C(x)$ itself rises to a maximum at $\xopt\approx0.34$ \cite{Boehmer2015,Reid2016} and upon further doping drops to a value that remains finite all the way to $x=1$. At low doping, superconductivity coexists with antiferromagnetism and orthorhombicity \cite{Avci2011, Wiesenmayer2011, Reid2016, Mallett2017}.

\KBa\ is special among the pnictides in that other phases have been reported in a narrow sliver of doping near $x\approx0.28$, separating the coexistence at low doping and the  superconducting phase at higher doping \cite{Avci2011, Boehmer2015, Hardy2016, Cho2016, Mallett2015}. Just above \Tc\ this sliver contains a tetragonal out-of-plane antiferromagnetic phase \cite{Boehmer2015, Mallett2015, Mallett2017} which coexists with superconductivity below \Tc. The superconducting phase in \KBa\ has its own unique attributes and affords unique opportunities that are not possible in other FeSCs where different phenomena occur in overlapping doping regimes. For example, in \KBa\ the coexistence regime is well below \Xopt. Moreover, the superconducting gap itself is nodeless below the highly doped regime, for which multigap superconductivity \cite{Malaeb2012,Hardy2016}, and the formation of gap anisotropy and nodes have been reported \cite{Goko2009, Mu2009,Malaeb2012,Xu2013, Cho2016}.

The effect of doping in \KBa\ is qualitatively different from other members of the \BaFeAs\ family \cite{Mallett2017}. Unlike the dopant Co, K is non-magnetic \cite{Athena2008}, and unlike the non-magnetic P, isovalent with As \cite{Hashimoto2012,Lamhot2015}, K adds holes. In addition, it is thought that \KBa\ is less disordered than other pnictides because the Ba sites hosting the K dopants are off the Fe-As planes \cite{Rotter2008, Rotter2010, Hardy2016, Mallett2017}
. All of this has motivated much research on superconductivity in \KBa\ \cite{Khasanov2009, Ohishi2012, Cho2014, Cho2016,Mallett2017,Hardy2016}, as well as on the structural \cite{Khan2014, Boehmer2015} and electronic \cite{Liu2014,Hardy2016} properties.

Here we report measurements of the absolute value of the penetration depth for currents flowing in the crystal a-b plane (\Lam) at low $T$ in high quality \KBa\ single crystals ranging from underdoped to overdoped. Frequently the measurement of \Lam\  \cite{Prozorov2006} is restricted to variations with temperature ($T$) \cite{Martin2009, Kim2014,Cho2016}. This provides information on the excitation spectrum rather than on the superfluid density itself ($\rho_s\equiv1/\lam^2$). Using MFM, we can measure the absolute value of \Lam\, and thus determine the superfluid density $\rho_s$ directly \cite{Prozorov2006,Gordon2010, Luan2011, Lamhot2015, Yagil2016}. The variation of $\rho_s(T=0)$ with doping is influenced by competition between superconductivity and other phases, as well as by changes in the band structure that can affect properties such as the effective mass  \cite{Hashimoto2012}. We also report pinning force measurements acquired by the manipulation of superconducting vortices   \cite{Auslaender2009, Yang2012, Shapira2015, Yagil2016}. Potentially this allows us to explore the impact of the structural and nematic phases at low doping  on vortex motion \cite{Avci2011, Wiesenmayer2011, Boehmer2015}.

Our measurements are local with the imaging resolution limit set by superconductivity itself to be on the order of \Lam. This allows us to go beyond sample-wide measurements \cite{Prozorov2006, Gordon2010, Cho2016, Mallett2017} and provide spatially resolved information.  For example, by obtaining \Lam\ and \Tc\ at the same location we can elucidate the relationship between these two fundamental quantities regardless of their variation across the sample \cite{Lamhot2015}.

\section{\label{sec:experiment}Experiment}
\subsection{Samples}
Our samples are single crystals grown by the self-flux method \cite{Liu2014,Cho2016} with Fe-As flux for samples with $x\leq0.55$ and K-As flux for higher levels of doping. The samples all have a surface area on the scale of $\approx0.25~\mathrm{mm}^2$ and a thickness of dozens of microns. The doping levels are $x=0.58\pm0.02,~0.52\pm0.01,~0.36\pm0.01,~0.34\pm0.01,~0.32\pm0.01,~0.24\pm0.01,~0.19\pm0.01$, spanning the superconducting dome. We determined these values by energy dispersive X-ray spectroscopy (EDS), which collects data from an area of $\approx1\times1$\um$^2$ at the actual scanned surface. The values listed above give the mean and the standard deviation from measurements at 10 different points across each sample. In addition to $x$, EDS gives the atomic composition, which was as expected [As ($37.6\%-42.5\%$), Fe ($38.1\%-41.0\%$)].

\subsection{Measurement}
\begin{figure}
\centering
\includegraphics[width=3.4in]{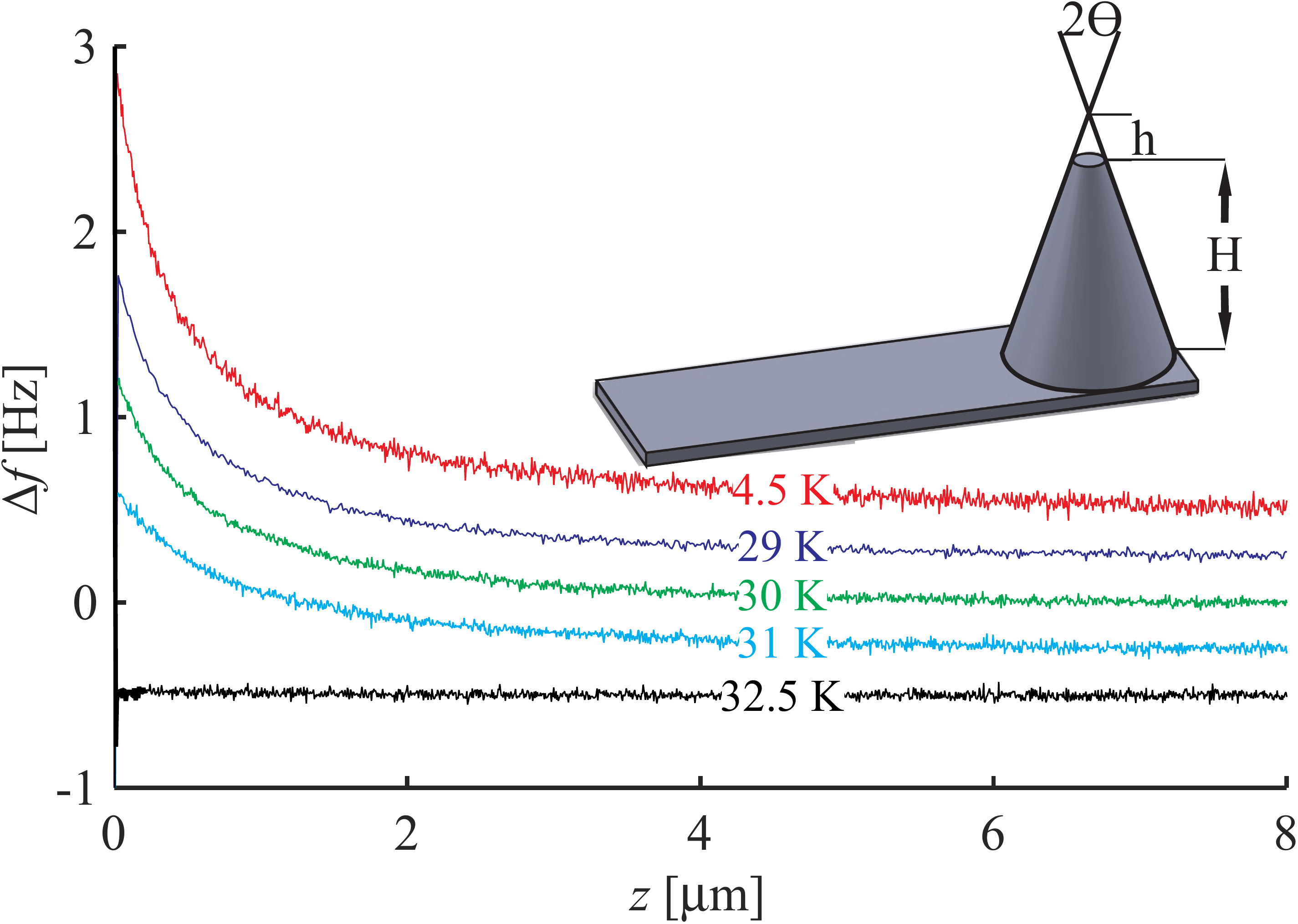}
	\caption{Touchdown curves as a function of $T$ for a $x=0.32$ sample showing how the increase of \Lam\ affects the repulsion of the tip from the surface. This sample was not used for extracting \Lam\ because it did not cleave well. All the curves here were acquired with the same tip at the same location and are offset by $0.25$~Hz for clarity. At T$=32.5$~K \Lam\ is too large for us to detect any Meissner response. Based on this and additional touchdown curves, $\tc=32.2\pm 0.2$~K. \textbf{Inset}: Schematic of an MFM tip. The truncated cone tip parameters are shown. $2\Theta$ is the cone angle, $H$ is the effective magnetic coating height, and $h$ is the truncation height.}
	\label{fig:Fig1}
\end{figure}
\begin{figure}
 \centering\includegraphics[width=3.4in]{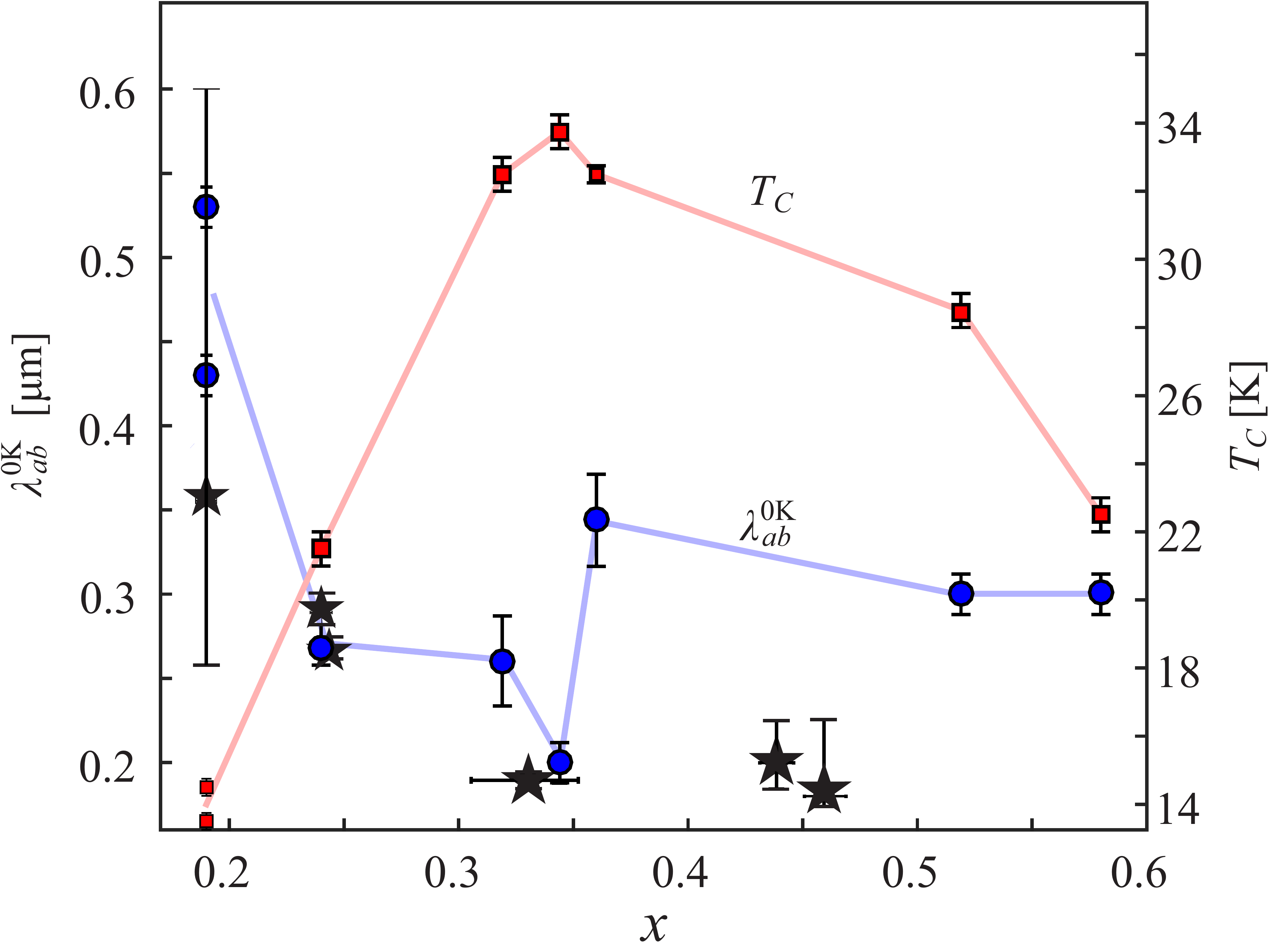}
\caption{The dependence of \Lamz\ (circles) and \Tc\ (squares) on doping $x$. \Lamz\ is extrapolated from $T=4.5$~K using data from Cho \etal \cite{Cho2016}. Stars are published values \cite{Mallett2015, Mallett2017} measured at $7$~K and extrapolated to $T=0$~K. The abrupt jump in \Lamz\ is clearly visible in our data at $x=0.36$, as is the decrease upon approaching \Xopt\ from the underdoped edge of the superconducting dome. For $x=0.19$ we show two values for \Lamz\ and \Tc, as explained in the text. The error bars for \Lamz\ represent $70\%$ confidence intervals. The error bars for \Tc\ represent temperature increments. Lines are guides to the eye.}
	\label{fig:Fig2}
\end{figure}
Prior to a measurement run we cleaved a sample to be scanned unless it already had a smooth ab-surface that showed no obvious signs of contamination. Thus we cleaved all samples except the $x=0.34$ sample. %
For the measurement we used frequency modulated MFM \cite{Albrecht1991} to determine the interaction between a sharp magnetic tip and a superconducting sample by tracking the frequency shift ($\Delta f$) of the resonant frequency of the cantilever holding the magnetic tip:
\begin{equation}\label{eq:dfdef}
\Delta f \approx C_{offset}-\frac{f_{0}}{2k} \frac{\partial F_{z}}{\partial z}.
\end{equation}
Here $z$ is the distance between the bottom of the MFM tip and the surface, $C_{offset}$ is an arbitrary constant offset, $f_{0}$ is the cantilever resonance frequency in free space, and $
k$ is its spring constant \cite{n:Tip}. $F_{z}$, a function of $\lam$ and $z$, is the $z$-component of the force between the tip and the sample. Equation~\ref{eq:dfdef} is an approximation for small oscillation amplitudes and $\Delta f\ll f_0$. $F_{z}$ also depends on the electric potential of the tip relative to the sample. When we tune it away from  the contact potential difference between the two, the MFM is sensitive to topography. When we tune it to cancel the contact potential difference, the only contribution is from magnetic forces \cite{Auslaender2009, Luan2011, Yang2012, Shapira2015, Lamhot2015, Yagil2016} for the range of $z$ we use for analysis here. 

Most of the results we report 
are from the Meissner repulsion of the tip from the sample, which we use to determine $\lam$. For this we acquire a touchdown curve: A measurement of $\Delta f(z)$ at a single point on the surface (e.g. \fig\ref{fig:Fig1}). Before such a measurement we field-cool the sample to control the density of superconducting vortices ($n_v$), which gives the magnetic field we report $B=\Phi_0n_v$, where $\Phi_0=hc/2e$ is the quantum of superconducting flux. To make sure that the only contribution to a touchdown is from the Meissner repulsion of the magnetic tip we use MFM imaging to locate a point which is at least $4$~\micro{m} from the nearest vortex, and is away from the sample edge or any other obvious defects.

Touchdown curves allow us to estimate \Tc: 
We define \Tc\ as the temperature where $\lam$ is too large to give detectable Meissner repulsion. We show an example in Fig.~\ref{fig:Fig1}. The disappearance of the Meissner repulsion results from the divergence of \Lam\ near \Tc\ \cite{Tinkham1996}. Based on our signal to noise ratio, our model and real tip parameters, we estimate that we can measure a Meissner response for $\lam \leq 10$~\micro{m}. Thus, our procedure gives lower bound on \Tc.

We extract $\lam$ from a touchdown curve by a fit 
that relies on a model of our tip. This model (the truncated cone model  \cite{Luan2010,Luan2011,Lamhot2015}) contains several parameters (cf. inset to Fig.~\ref{fig:Fig1}). We determined some of them (the cone angle $2\Theta$ and the truncation height $h$) by scanning electron microscopy (SEM). Additional tip parameters (the cone effective magnetic height $H$ and an overall prefactor $A$) are more difficult to determine as they are affected by the magnetic domain structure of the tip, which we have not measured directly. We determine these last parameters together with $\lam$ and $C_{offset}$ in a fit process, as described previously \cite{Lamhot2015}. %
Once we have a value for \Lam\ we obtain the $T=0$~K value  (\Lamz) by extrapolation using published data on the temperature dependence \cite{Cho2016}, which changes \Lam\ by $\lesssim 50$~nm for $x=0.19$ and $\lesssim 10$~nm for $0.24\leq x\leq0.58$. The values we report in Fig.~\ref{fig:Fig2} for \Lamz\ are an average over several points in each sample. At each point we average over multiple touchdown curves.

In addition to measuring 
the Meissner response, we also imaged and manipulated superconducting vortices. Vortex motion and the mapping of vortex positions can give information on structure and the defect landscape \cite{Straver08, Auslaender2009,Yang2012, Zhang2015, Shapira2015}. For this we utilize the interaction between the magnetic MFM tip and the currents circulating the core of a vortex \cite{Straver08, Auslaender2009, Shapira2015, Lamhot2015, Yagil2016}. After field-cooling ($1~\mathrm{G}\lesssim|B|\lesssim3~\mathrm{G}$), we imaged the magnetic landscape with the tip far enough to leave the vortices unperturbed (surveillance scanning). For manipulation we brought the tip close enough to the surface to drag or to push vortices out of their pinning sites \cite{Auslaender2009,Zhang2015,Shapira2015,Yagil2016}.

\section{\label{sec:results}Results}

\subsection{Local diamagnetic response}
\begin{figure}
\centering\includegraphics[width=3.4in]{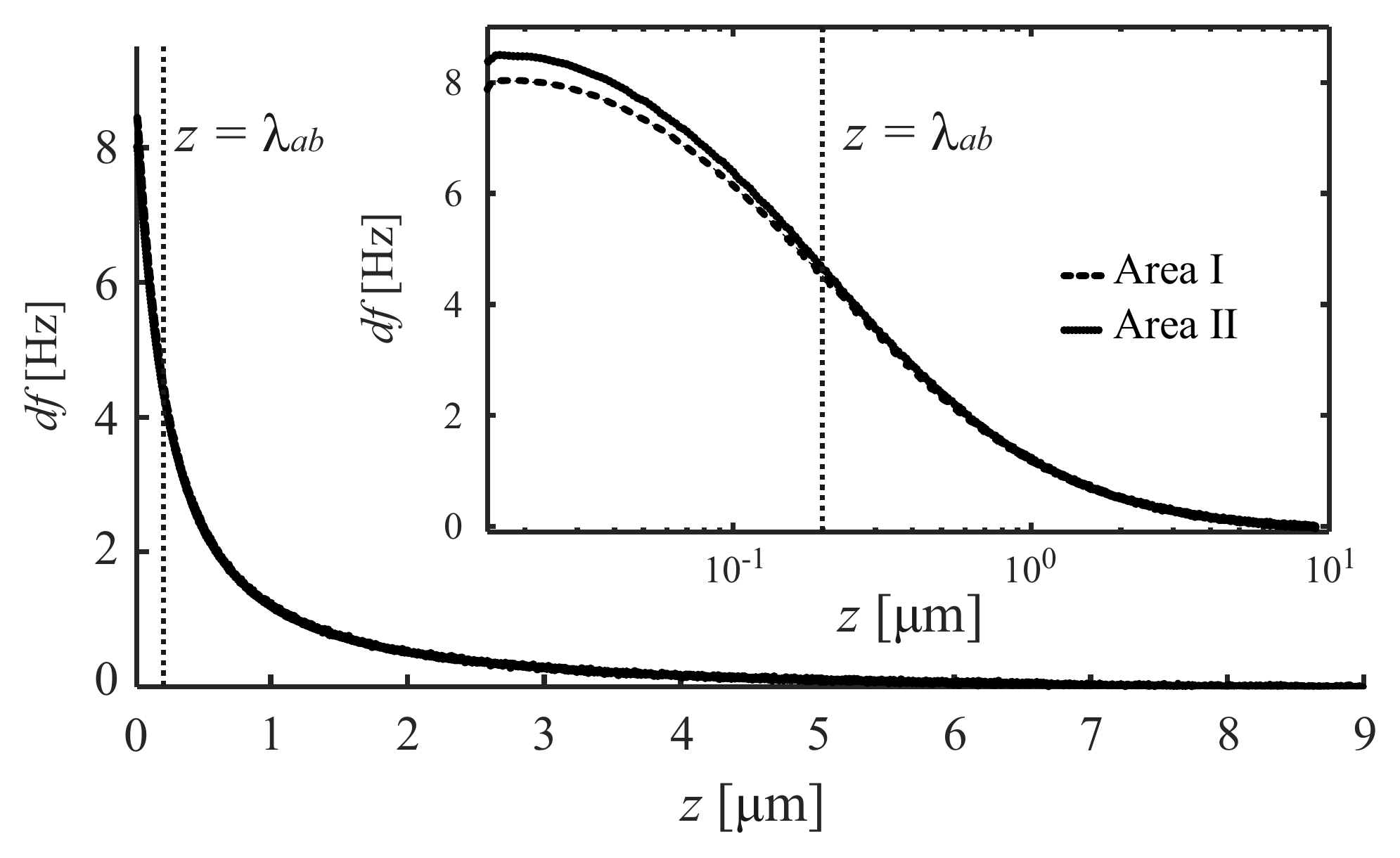}
\caption{Touchdown curves taken at points $\approx200$~\um\ apart on a $x=0.34$ sample during the same cool-down at $T=4.6$~K. Clearly the curves are very similar. Fitting gives $\lam=200\pm30$~nm. Vertical line represents $z=\lam$. For fitting we use the $z\geq2\lambda_{ab}$ part of the data. $\mathbf{Inset:}$ same touchdowns presented with $z$ on a logarithmic scale showing the similarity for $z\geq2\lambda_{ab}$.}
\label{fig:Fig3}
\end{figure}

Figure~\ref{fig:Fig2} shows our main results: 
The dependence of \Lamz\ and \Tc\ on doping. In all of the samples except at the lowest doping ($x=0.19$) \Lam\ and \Tc\ were uniform with the scatter for \Lam\ below $30$~nm. This uniformity is reflected in the touchdown curves themselves. For example, Fig.~\ref{fig:Fig3} shows two touchdown curves taken $\approx200$~\micro{m} apart on a  $x=0.34$ sample. Clearly the curves are very similar,  attesting to the uniformity of \Lam\ in this sample.

We account for the scatter  of \Lam\ 
and \Tc\ at $x=0.19$ by showing two separate results for data acquired at different points during the same cool-down (cf. Fig.~\ref{fig:Fig2}). This is likely a consequence of the strong dependence of \Lamz\ and \Tc\ on doping at low $x$ and indicates doping variations across the sample. This matches both our EDS results, where we see variations of $x$ on the scale of $\pm0.01$, and the known tendency of K to be distributed inhomogeneously in \KBa\ \cite{Park2009,Ohgushi2012}. Similar scatter in very underdoped samples has been observed in underdoped \PBa\ \cite{Lamhot2015}. The scatter shows one of the advantages of our local probe: Instead of extracting an average value for a whole sample, we can extract different values from different parts of the sample.

\sloppy
The dependence of \Tc\ 
on $x$ shows the dome typical to the FeSC \cite{Gordon2010, Avci2011, Ohgushi2012, Liu2014, Boehmer2015, Cho2016}. As expected, \Tc\ increases sharply when $x$ is increased from the underdoped side towards \Xopt, and decreases slowly when $x$ is increased further towards the overdoped side. The \Tc\ values we obtain are lower than previously reported in sample-wide measurements on similar materials \cite{Liu2014,Boehmer2015,Cho2016} and the variation around \Xopt\ is sharper, as expected from our technique, which gives a lower-bound.
We have observed similar behavior of $\tc(x)$ in \PBa\ \cite{Lamhot2015}, which is reminiscent of the saturation of diamagnetic signal rather than its onset in sample-wide measurements \cite{Hashimoto2012}.

The overall dependence of \Lamz\ 
on $x$ is reminiscent of the dependence in \CoBa\ \cite{Gordon2010, Luan2011}, in which there is a sharp drop from the underdoped edge of the superconducting dome followed by a shallow minimum around \Xopt\ and a leveling off for $x>\xopt$. The sharp drop in \Lam\ on the underdoped side has also been reported in \PBa\ \cite{Lamhot2015}. This kind of behavior can be attributed to the competition of superconductivity with a spin-density-wave phase in the coexistence region of phase diagram \cite{Gordon2010, Luan2011, Lamhot2015}.

\begin{figure}
\centering\includegraphics[width=3.4in]{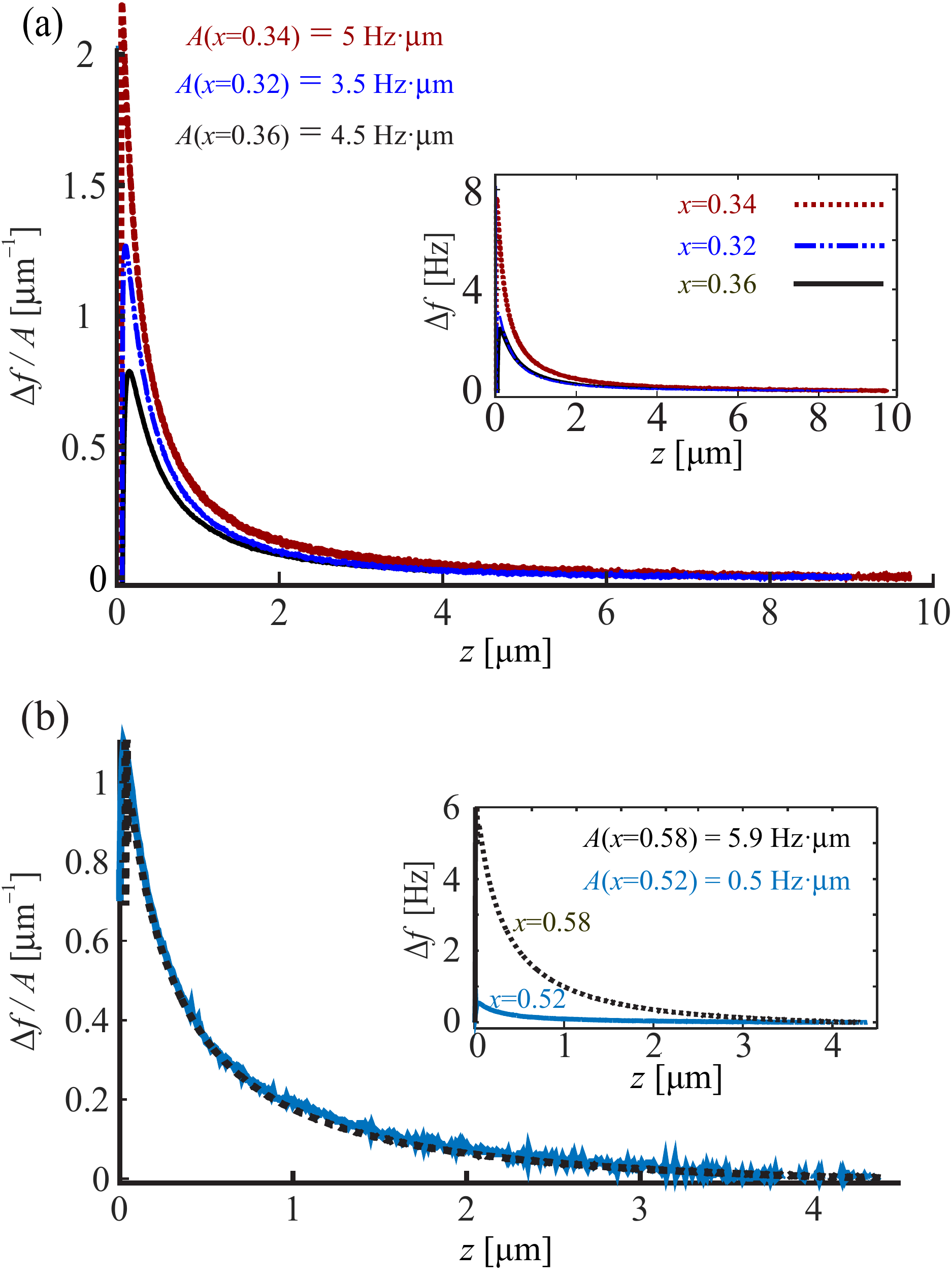}
\caption{
\textbf{(a)} Normalized touchdown curves measured at $T=4.5$~K for $x=0.34,~0.32,~0.36$. $x=0.32,~0.36$ were measured in the same cool-down with the same tip. $x=0.34$ was measured with a different tip in a different cool-down. Fitting to the curves gives  $\lam \approx200\pm30,~260\pm30,~340\pm50$~nm for $x=0.34,~0.32,~0.36$. \textbf{Inset}: The same curves before normalization. %
\textbf{(b)} Normalized touchdown curves for different samples ($x = 0.58,~0.52$) acquired with different tips. Both give $\lam \approx300 \pm 35$~nm at $T=4.5$~K. The $x=0.52$ curve is offset by $20$~nm to emphasize the similarity to the $x=0.58$ curve. \textbf{Inset}: The same curves before normalization.
}
\label{fig:Fig4}
\end{figure}

The most surprising behavior we observe 
in Fig.~\ref{fig:Fig2} is an abrupt jump of \Lamz\ when $x$ is slightly increased from \Xopt. This observation is based on measurements in three samples with $x=0.32$, $0.34$ and $0.36$. To help rule out an artifact of using different tips we show full touchdown curves in \Fig{fig:Fig4}(a). To compare curves that were acquired with different tips we normalized the raw data by the prefactor $A$, the fit parameter which is proportional to the magnetization of the tip. We show in \Fig{fig:Fig4}(b) that the difference between the curves is due primarily to the variation of \Lam\ rather than the tip parameters by comparing normalized plots acquired with different tips but with the fit procedure yielding similar values of \Lam.

\subsection{Imaging and manipulation of vortices}
\begin{figure}
\centering
\vspace*{3mm}
\includegraphics[width=3.4in]{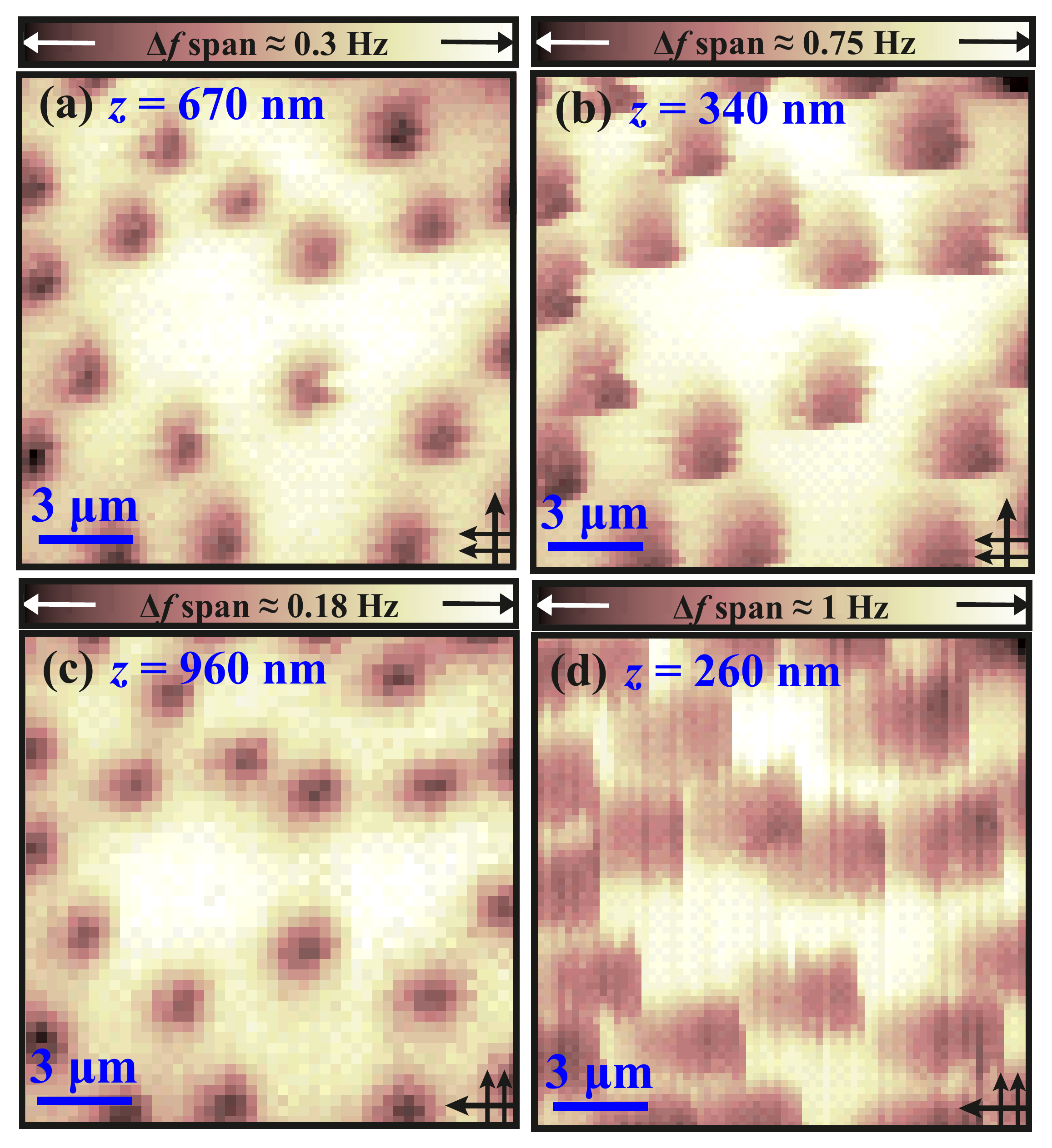}
\caption{Imaging and manipulating vortices at $x=0.58$ for $ T= 4.3$~K with $B\approx 1.8$~G. The scans show vortex motion which depends on the scan direction (indicated by arrows -- the fast direction, in which we move the tip back-and-forth,  by two parallel arrows; the slow direction, in which we increment the tip after one back-and-forth period, by a single long arrow). %
	\textbf{(a)} $z=670$~nm. %
	\textbf{(b)} $z=340$~nm. %
	\textbf{(c)} $z=960$~nm. %
	\textbf{(d)} $z=260$~nm.}
	\label{fig:Fig5}
\end{figure}
\begin{figure*}
	\centering
	\includegraphics[width=5.2in]{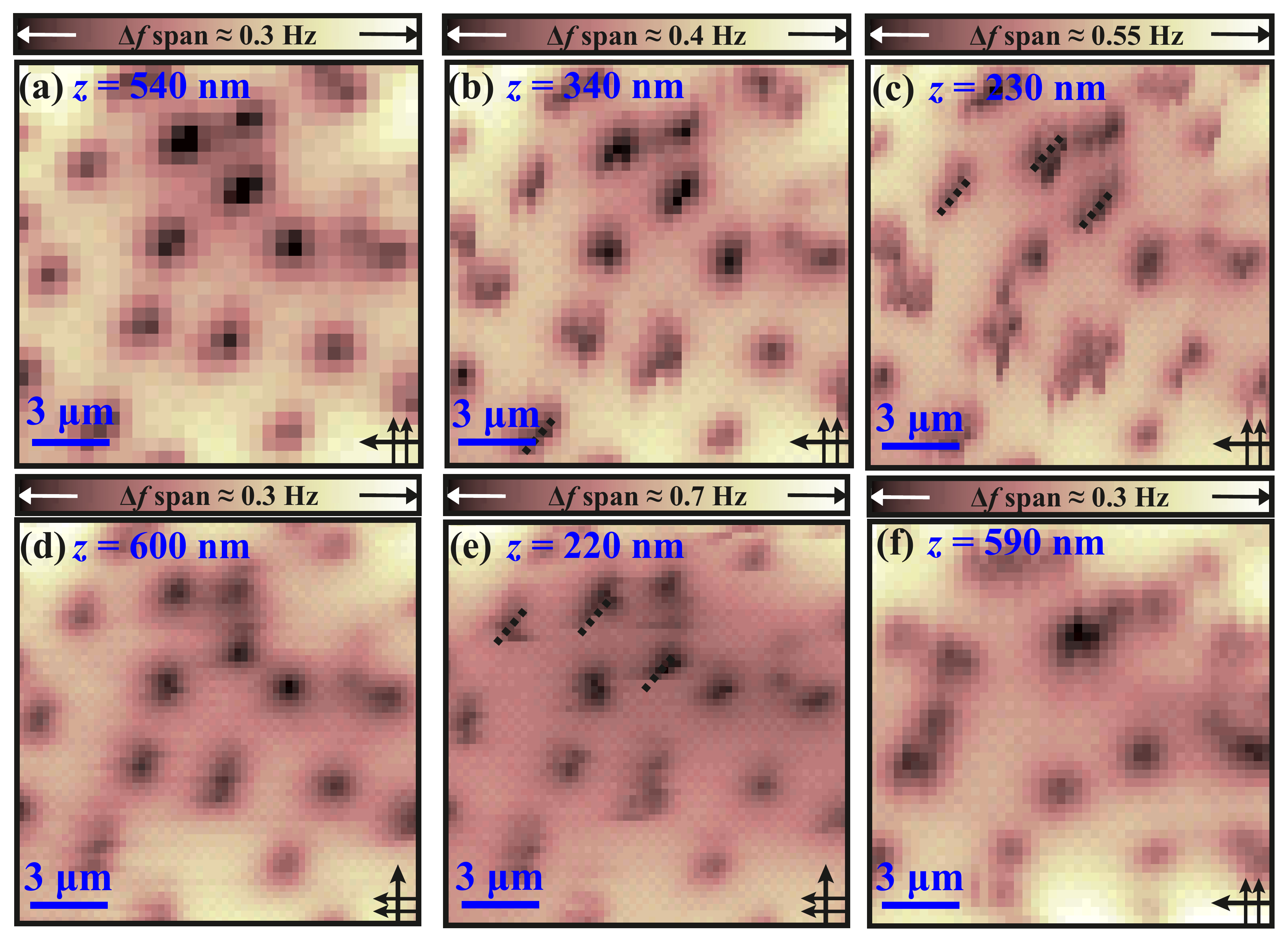}
\caption[%
Vortex imaging and manipulation at $x=0.19$]{Imaging and manipulating vortices at $x=0.19$ for $T= 4.34$~K with $B \approx1.5$~G. The scan directions are indicated by arrows as explained in Fig.~\ref{fig:Fig5}.  The scans are ordered chronologically. Dashed lines in (c),(e) are guides to the eye and highlight vortex motion. %
\textbf{(a)} Low resolution scan before manipulating vortices ($z=540$~nm). %
\textbf{(b),(c)} Manipulation scans with the slow scan direction pointing left [$z=340$~nm in (b), $z=230$~nm in (c)]. %
\textbf{(d),(e)} Scans with the slow scan direction pointing up [$z=600$~nm in (d), $z=220$~nm in (e)].  %
\textbf{(f)} Scan with $z=590$~nm after several scans with very low $z$ and significant vortex motion (not shown).}
	\label{fig:Fig6}
\end{figure*}
%

Overall our conclusion from imaging vortex positions is that the disorder level in all samples is low -- vortices did not cluster, an indication for the absence of strong pinning sites which overwhelm vortex-vortex interactions when vortices freeze in place during a cool-down \cite{Luan2010}.
We also probed samples by dragging vortices. 
For example, anisotropic vortex motion can be an indication for the presence of twin boundaries \cite{Kalisky2011, Lamhot2015, Yagil2016}, nematic order, or other broken symmetries. To achieve controlled vortex motion we cooled samples in a field aligned with the magnetization of the tip. This gives tip-vortex attraction and vortices that appear as dark spots (\Figs{fig:Fig5},~\ref{fig:Fig6}). We were able to move vortices in three of the samples ($x=0.19$, $0.52$, $0.58$) and studied them in detail in two where vortex motion was substantial and qualitatively different ($x=0.19,~0.58$). %
The pinning forces measured for the manipulated samples were much smaller than reported for \PBa\ \cite{Yagil2016} and \CoBa\  \cite{Luan2010}. This is an indication of weak vortex pinning \cite{Auslaender2009, Lamhot2015, Yagil2016} \cite{n:pinning_forces}.

Figure~\ref{fig:Fig5} shows both surveillance 
scans for the $x=0.58$ sample [(a),(c)] as well as manipulation scans [(b),(d)]. Tip-induced motion for different vortices started at $670~\mathrm{nm}\geq z\geq340~\mathrm{nm}$, which suggests  that the range of pinning force in this sample was $1.7~\mathrm{pN}\lesssim \fp\lesssim2.6~\mathrm{pN}$. For such an estimate we perform a sequence of surveillance scans, each one closer to the sample. We estimate \Fp\ for a particular vortex from the maximum of the lateral force\cite{n:estimatingF} (\Flatmax) that  we apply in the first scan for which we see it move. The motion of vortices did not show an obvious preferred orientation -- they tracked the slow axis of raster pattern in perpendicular scan-orientations, as in \fig\ref{fig:Fig5}(b),(d). The lack of a clear preferred axis is consistent with the tetragonal symmetry ($C_4$) known to exist in  overdoped \KBa\ \cite{Liu2014,Boehmer2015}. The way vortices crept along the slow axis is reminiscent of the behavior in slightly overdoped \YBCO\ (clean samples with low anisotropy) \cite{Auslaender2009}. Indeed, as in \YBCO, all of the vortices jumped back towards their original pinning site once the tip was far enough away  [cf. \Figs{fig:Fig5}(a),(c)].

Vortex motion was different in the $x=0.19$ sample. 
Figure~\ref{fig:Fig6}(a) shows unperturbed vortices at $z=540$~nm. Next is a scan for $z=340$~nm [\Fig{fig:Fig6}(b)] with significant vortex motion. Our estimate of \Flatmax\ \cite{n:estimatingF} suggests that the pinning force in this sample was $1.6\lesssim\fp\lesssim2.0~\mathrm{pN}$. Reducing $z$ further increased the tip-vortex force and allowed us to move vortices even more. This is shown in \Figs{fig:Fig6}(c),(e). Close inspection of these scans suggests a preferred direction for vortex motion (shown by dashed lines), that is independent of the scan orientation. This is consistent with broken $C_4$ symmetry and the existence of orthorhombic domains and the twin boundaries that separate them.  Twin boundaries have been observed previously at this doping \cite{Yang2012,Boehmer2015} -- their presence is an indication that this sample is in the coexistence regime. A scan performed from a higher scan height between these two scans [\Fig{fig:Fig6}(d)] shows that in this sample vortices returned to their original positions after mild perturbation.

We subjected the vortices in the $x=0.19$ sample 
to even stronger dragging forces by scanning at $z=100$~nm, where the tip exerts a force as large as $\flatmax\approx3$~pN \cite{n:estimatingF}. After this strong manipulation we scanned with a larger $z$ (to reduce \Flatmax) to determine the ultimate positions of the vortices. As \Fig{fig:Fig6}(f) shows, $\flatmax\approx3$~pN was sufficient to pull vortices far from their original pinning sites. The position changes of vortices under strong perturbation, and the scale of the forces applied, lead us to conclude that if there are sites of strong pinning, they are rare. This further attests to the high quality of the samples.

\section{Discussion}
Our values for \Lam\ are in agreement with estimates from  infrared reflectivity \cite{Mallett2015, Mallett2017} (stars in \Fig{fig:Fig2}) only for $x\leq\xopt$. For $x>\xopt$ our values are higher, perhaps because in \KBa\  this is a  strongly-coupled regime \cite{Shan2012, Hardy2016}, where reflectivity provides a lower bound on \Lam\  \cite{Hirsch1992, Kogan2009}.

Our most surprising result 
is the abrupt increase of \Lamz\ when $x$ is tuned up from \Xopt. The only FeSC where anything remotely similar has been observed is \PBa, where \Lamz\ has a peak at \Xopt\ \cite{Hashimoto2012, Lamhot2015} that coincides with the upper boundary of the coexistence regime. It is possible that the increase that we see at $x=0.36$ is part of a peak that therefore also exists in \KBa, but until additional samples are measured, especially for  $0.35\lesssim{x}\lesssim0.5$, it is impossible to be certain.

If the sharp increase of \Lamz\  
is indeed part of a peak, then this peak exists well beyond the reported coexistence range $x\lesssim0.28$ \cite{Boehmer2015}, and thus may hint at the presence of another phase. But, unless magnetic phases are detected near optimal doping, a micro-emulsion mechanism of the type that was invoked to explain the peak in \Lamz\ in \PBa\ \cite{Chowdhury2015} probably does not play a role. A new phase could be the reason masses renormalize and, through that, the reason for \Lamz\ to increase \cite{Wang2016}. In fact, measurements of the Hall coefficient suggest an increase of the ratio between the hole and electron effective masses \cite{Ohgushi2012}. This has been interpreted as a consequence of the creation of a coherent electronic state in which holes interact via bosons. This boson-hole interaction \cite{Ohgushi2012} may also influence the coupling of the cooper-pairs, as measurements of the specific heat \cite{Hardy2016} imply. Interestingly, scanning tunneling spectroscopy (STS) experiments have reported bosonic modes that have a relationship with the superconducting order parameter \cite{Shan2012}, and are an indication of strong coupling.

A tantalizing explanation  
for the observed increase in \Lamz, that may also explain the boson-hole interaction and the mass renormalization reported previously \cite{Ohgushi2012}, is the existence of quantum critical point (QCP). %
The peaked \Lamz\ at \Xopt\ in \PBa\ has been associated with such a QCP  \cite{Hashimoto2012,Levchenko2013,Fernandes2013,Nomoto2013,Shibauchi2014}, although this view is not uncontested \cite{Chowdhury2013,Chowdhury2015}. If our observed increase of \Lamz\ is indeed a result of a QCP this implies that the nodal gap structure of \PBa\ is not a consequence of the quantum critical behavior, as the gap in \KBa\ is nodeless near \Xopt\ \cite{Shibauchi2014,Cho2016}.
That \CoBa, the gap of which is also nodeless near \Xopt,  does not show this behavior is most likely  because it is in the dirty limit \cite{Gordon2010b}. %
On the other hand, it is believed that magnetic order is crucial for the peaked behavior of \Lamz\ in \PBa, but this order is absent near \Xopt\ in \KBa. %

\begin{acknowledgments} 
We would like to thank A. Chubukov, B. Kalisky, A. Kanigel, I. Kapon, and M. Khodas for discussions, A. Brenner for help with EDS, as well as the Micro Nano Fabrication Unit at the Technion. This work was supported by the Israel Science Foundation (grant no. 1897/14). Work at Ames was supported by the U.S. Department of Energy (DOE), Office of Science, Basic Energy Sciences, Materials Science and Engineering Division. Ames Laboratory is operated for the U.S. DOE by Iowa State University under contract \# DE-AC02-07CH11358.
\end{acknowledgments}

\bibliographystyle{apsrev4-1}

\begin{thebibliography}{60}%
\makeatletter
\providecommand \@ifxundefined [1]{%
 \@ifx{#1\undefined}
}%
\providecommand \@ifnum [1]{%
 \ifnum #1\expandafter \@firstoftwo
 \else \expandafter \@secondoftwo
 \fi
}%
\providecommand \@ifx [1]{%
 \ifx #1\expandafter \@firstoftwo
 \else \expandafter \@secondoftwo
 \fi
}%
\providecommand \natexlab [1]{#1}%
\providecommand \enquote  [1]{``#1''}%
\providecommand \bibnamefont  [1]{#1}%
\providecommand \bibfnamefont [1]{#1}%
\providecommand \citenamefont [1]{#1}%
\providecommand \href@noop [0]{\@secondoftwo}%
\providecommand \href [0]{\begingroup \@sanitize@url \@href}%
\providecommand \@href[1]{\@@startlink{#1}\@@href}%
\providecommand \@@href[1]{\endgroup#1\@@endlink}%
\providecommand \@sanitize@url [0]{\catcode `\\12\catcode `\$12\catcode
  `\&12\catcode `\#12\catcode `\^12\catcode `\_12\catcode `\%12\relax}%
\providecommand \@@startlink[1]{}%
\providecommand \@@endlink[0]{}%
\providecommand \url  [0]{\begingroup\@sanitize@url \@url }%
\providecommand \@url [1]{\endgroup\@href {#1}{\urlprefix }}%
\providecommand \urlprefix  [0]{URL }%
\providecommand \Eprint [0]{\href }%
\providecommand \doibase [0]{http://dx.doi.org/}%
\providecommand \selectlanguage [0]{\@gobble}%
\providecommand \bibinfo  [0]{\@secondoftwo}%
\providecommand \bibfield  [0]{\@secondoftwo}%
\providecommand \translation [1]{[#1]}%
\providecommand \BibitemOpen [0]{}%
\providecommand \bibitemStop [0]{}%
\providecommand \bibitemNoStop [0]{.\EOS\space}%
\providecommand \EOS [0]{\spacefactor3000\relax}%
\providecommand \BibitemShut  [1]{\csname bibitem#1\endcsname}%
\let\auto@bib@innerbib\@empty
\bibitem [{\citenamefont {{Lan Luan}}\ \emph {et~al.}(2011)\citenamefont {{Lan
  Luan}}, \citenamefont {Lippman}, \citenamefont {Hicks}, \citenamefont {Bert},
  \citenamefont {Auslaender}, \citenamefont {Chu}, \citenamefont {Analytis},
  \citenamefont {Fisher},\ and\ \citenamefont {Moler}}]{Luan2011}%
  \BibitemOpen
  \bibfield  {author} {\bibinfo {author} {\bibnamefont {{Lan Luan}}}, \bibinfo
  {author} {\bibfnamefont {T.~M.}\ \bibnamefont {Lippman}}, \bibinfo {author}
  {\bibfnamefont {C.~W.}\ \bibnamefont {Hicks}}, \bibinfo {author}
  {\bibfnamefont {J.~A.}\ \bibnamefont {Bert}}, \bibinfo {author}
  {\bibfnamefont {O.~M.}\ \bibnamefont {Auslaender}}, \bibinfo {author}
  {\bibfnamefont {J.-H.}\ \bibnamefont {Chu}}, \bibinfo {author} {\bibfnamefont
  {J.~G.}\ \bibnamefont {Analytis}}, \bibinfo {author} {\bibfnamefont {I.~R.}\
  \bibnamefont {Fisher}}, \ and\ \bibinfo {author} {\bibfnamefont {K.~A.}\
  \bibnamefont {Moler}},\ }\href {\doibase 10.1103/PhysRevLett.106.067001}
  {\bibfield  {journal} {\bibinfo  {journal} {Phys. Rev. Lett.}\ }\textbf
  {\bibinfo {volume} {106}},\ \bibinfo {pages} {067001} (\bibinfo {year}
  {2011})}\BibitemShut {NoStop}%
\bibitem [{\citenamefont {Avci}\ \emph {et~al.}(2011)\citenamefont {Avci},
  \citenamefont {Chmaissem}, \citenamefont {Goremychkin}, \citenamefont
  {Rosenkranz}, \citenamefont {Castellan}, \citenamefont {Chung}, \citenamefont
  {Todorov}, \citenamefont {Schlueter}, \citenamefont {Claus}, \citenamefont
  {Kanatzidis}, \citenamefont {Daoud-Aladine}, \citenamefont {Khalyavin},\ and\
  \citenamefont {Osborn}}]{Avci2011}%
  \BibitemOpen
  \bibfield  {author} {\bibinfo {author} {\bibfnamefont {S.}~\bibnamefont
  {Avci}}, \bibinfo {author} {\bibfnamefont {O.}~\bibnamefont {Chmaissem}},
  \bibinfo {author} {\bibfnamefont {E.~A.}\ \bibnamefont {Goremychkin}},
  \bibinfo {author} {\bibfnamefont {S.}~\bibnamefont {Rosenkranz}}, \bibinfo
  {author} {\bibfnamefont {J.-P.}\ \bibnamefont {Castellan}}, \bibinfo {author}
  {\bibfnamefont {D.~Y.}\ \bibnamefont {Chung}}, \bibinfo {author}
  {\bibfnamefont {I.~S.}\ \bibnamefont {Todorov}}, \bibinfo {author}
  {\bibfnamefont {J.~A.}\ \bibnamefont {Schlueter}}, \bibinfo {author}
  {\bibfnamefont {H.}~\bibnamefont {Claus}}, \bibinfo {author} {\bibfnamefont
  {M.~G.}\ \bibnamefont {Kanatzidis}}, \bibinfo {author} {\bibfnamefont
  {A.}~\bibnamefont {Daoud-Aladine}}, \bibinfo {author} {\bibfnamefont
  {D.}~\bibnamefont {Khalyavin}}, \ and\ \bibinfo {author} {\bibfnamefont
  {R.}~\bibnamefont {Osborn}},\ }\href {\doibase 10.1103/PhysRevB.83.172503}
  {\bibfield  {journal} {\bibinfo  {journal} {Phys. Rev. B}\ }\textbf {\bibinfo
  {volume} {83}},\ \bibinfo {pages} {172503} (\bibinfo {year}
  {2011})}\BibitemShut {NoStop}%
\bibitem [{\citenamefont {B\"{o}hmer}\ \emph {et~al.}(2015)\citenamefont
  {B\"{o}hmer}, \citenamefont {Hardy}, \citenamefont {Wang}, \citenamefont
  {Wolf}, \citenamefont {Schweiss},\ and\ \citenamefont
  {Meingast}}]{Boehmer2015}%
  \BibitemOpen
  \bibfield  {author} {\bibinfo {author} {\bibfnamefont {A.~E.}\ \bibnamefont
  {B\"{o}hmer}}, \bibinfo {author} {\bibfnamefont {F.}~\bibnamefont {Hardy}},
  \bibinfo {author} {\bibfnamefont {L.}~\bibnamefont {Wang}}, \bibinfo {author}
  {\bibfnamefont {T.}~\bibnamefont {Wolf}}, \bibinfo {author} {\bibfnamefont
  {P.}~\bibnamefont {Schweiss}}, \ and\ \bibinfo {author} {\bibfnamefont
  {C.}~\bibnamefont {Meingast}},\ }\href
  {http://www.nature.com/ncomms/2015/150731/ncomms8911/full/ncomms8911.html}
  {\bibfield  {journal} {\bibinfo  {journal} {Nat. Commun.}\ }\textbf {\bibinfo
  {volume} {6}},\ \bibinfo {pages} {7911} (\bibinfo {year} {2015})}\BibitemShut
  {NoStop}%
\bibitem [{\citenamefont {Lamhot}\ \emph {et~al.}(2015)\citenamefont {Lamhot},
  \citenamefont {Yagil}, \citenamefont {Shapira}, \citenamefont {Kasahara},
  \citenamefont {Watashige}, \citenamefont {Shibauchi}, \citenamefont
  {Matsuda},\ and\ \citenamefont {Auslaender}}]{Lamhot2015}%
  \BibitemOpen
  \bibfield  {author} {\bibinfo {author} {\bibfnamefont {Y.}~\bibnamefont
  {Lamhot}}, \bibinfo {author} {\bibfnamefont {A.}~\bibnamefont {Yagil}},
  \bibinfo {author} {\bibfnamefont {N.}~\bibnamefont {Shapira}}, \bibinfo
  {author} {\bibfnamefont {S.}~\bibnamefont {Kasahara}}, \bibinfo {author}
  {\bibfnamefont {T.}~\bibnamefont {Watashige}}, \bibinfo {author}
  {\bibfnamefont {T.}~\bibnamefont {Shibauchi}}, \bibinfo {author}
  {\bibfnamefont {Y.}~\bibnamefont {Matsuda}}, \ and\ \bibinfo {author}
  {\bibfnamefont {O.~M.}\ \bibnamefont {Auslaender}},\ }\href {\doibase
  10.1103/PhysRevB.91.060504} {\bibfield  {journal} {\bibinfo  {journal} {Phys.
  Rev. B}\ }\textbf {\bibinfo {volume} {91}},\ \bibinfo {pages} {060504}
  (\bibinfo {year} {2015})}\BibitemShut {NoStop}%
\bibitem [{\citenamefont {Reid}\ \emph {et~al.}(2016)\citenamefont {Reid},
  \citenamefont {Tanatar}, \citenamefont {Luo}, \citenamefont {Shakeripour},
  \citenamefont {de~Cotret}, \citenamefont {Juneau-Fecteau}, \citenamefont
  {Chang}, \citenamefont {Shen}, \citenamefont {Wen}, \citenamefont {Kim} \emph
  {et~al.}}]{Reid2016}%
  \BibitemOpen
  \bibfield  {author} {\bibinfo {author} {\bibfnamefont {J.~P.}\ \bibnamefont
  {Reid}}, \bibinfo {author} {\bibfnamefont {M.~A.}\ \bibnamefont {Tanatar}},
  \bibinfo {author} {\bibfnamefont {X.~G.}\ \bibnamefont {Luo}}, \bibinfo
  {author} {\bibfnamefont {H.}~\bibnamefont {Shakeripour}}, \bibinfo {author}
  {\bibfnamefont {S.~R.}\ \bibnamefont {de~Cotret}}, \bibinfo {author}
  {\bibfnamefont {A.}~\bibnamefont {Juneau-Fecteau}}, \bibinfo {author}
  {\bibfnamefont {J.}~\bibnamefont {Chang}}, \bibinfo {author} {\bibfnamefont
  {B.}~\bibnamefont {Shen}}, \bibinfo {author} {\bibfnamefont {H.~H.}\
  \bibnamefont {Wen}}, \bibinfo {author} {\bibfnamefont {H.}~\bibnamefont
  {Kim}},  \emph {et~al.},\ }\href
  {http://link.aps.org/doi/10.1103/PhysRevB.93.214519} {\bibfield  {journal}
  {\bibinfo  {journal} {Phys. Rev. B}\ }\textbf {\bibinfo {volume} {93}},\
  \bibinfo {pages} {214519} (\bibinfo {year} {2016})}\BibitemShut {NoStop}%
\bibitem [{\citenamefont {Hashimoto}\ \emph {et~al.}(2012)\citenamefont
  {Hashimoto}, \citenamefont {Cho}, \citenamefont {Shibauchi}, \citenamefont
  {Kasahara}, \citenamefont {Mizukami}, \citenamefont {Katsumata},
  \citenamefont {Tsuruhara}, \citenamefont {Terashima}, \citenamefont {Ikeda},
  \citenamefont {Tanatar}, \citenamefont {Kitano}, \citenamefont {Salovich},
  \citenamefont {Giannetta}, \citenamefont {Walmsley}, \citenamefont
  {Carrington}, \citenamefont {Prozorov},\ and\ \citenamefont
  {Matsuda}}]{Hashimoto2012}%
  \BibitemOpen
  \bibfield  {author} {\bibinfo {author} {\bibfnamefont {K.}~\bibnamefont
  {Hashimoto}}, \bibinfo {author} {\bibfnamefont {K.}~\bibnamefont {Cho}},
  \bibinfo {author} {\bibfnamefont {T.}~\bibnamefont {Shibauchi}}, \bibinfo
  {author} {\bibfnamefont {S.}~\bibnamefont {Kasahara}}, \bibinfo {author}
  {\bibfnamefont {Y.}~\bibnamefont {Mizukami}}, \bibinfo {author}
  {\bibfnamefont {R.}~\bibnamefont {Katsumata}}, \bibinfo {author}
  {\bibfnamefont {Y.}~\bibnamefont {Tsuruhara}}, \bibinfo {author}
  {\bibfnamefont {T.}~\bibnamefont {Terashima}}, \bibinfo {author}
  {\bibfnamefont {H.}~\bibnamefont {Ikeda}}, \bibinfo {author} {\bibfnamefont
  {M.~A.}\ \bibnamefont {Tanatar}}, \bibinfo {author} {\bibfnamefont
  {H.}~\bibnamefont {Kitano}}, \bibinfo {author} {\bibfnamefont
  {N.}~\bibnamefont {Salovich}}, \bibinfo {author} {\bibfnamefont {R.~W.}\
  \bibnamefont {Giannetta}}, \bibinfo {author} {\bibfnamefont {P.}~\bibnamefont
  {Walmsley}}, \bibinfo {author} {\bibfnamefont {A.}~\bibnamefont
  {Carrington}}, \bibinfo {author} {\bibfnamefont {R.}~\bibnamefont
  {Prozorov}}, \ and\ \bibinfo {author} {\bibfnamefont {Y.}~\bibnamefont
  {Matsuda}},\ }\href {\doibase 10.1126/science.1219821} {\bibfield  {journal}
  {\bibinfo  {journal} {Science}\ }\textbf {\bibinfo {volume} {336}},\ \bibinfo
  {pages} {1554} (\bibinfo {year} {2012})}\BibitemShut {NoStop}%
\bibitem [{\citenamefont {Yagil}\ \emph {et~al.}(2016)\citenamefont {Yagil},
  \citenamefont {Lamhot}, \citenamefont {Almoalem}, \citenamefont {Kasahara},
  \citenamefont {Watashige}, \citenamefont {Shibauchi}, \citenamefont
  {Matsuda},\ and\ \citenamefont {Auslaender}}]{Yagil2016}%
  \BibitemOpen
  \bibfield  {author} {\bibinfo {author} {\bibfnamefont {A.}~\bibnamefont
  {Yagil}}, \bibinfo {author} {\bibfnamefont {Y.}~\bibnamefont {Lamhot}},
  \bibinfo {author} {\bibfnamefont {A.}~\bibnamefont {Almoalem}}, \bibinfo
  {author} {\bibfnamefont {S.}~\bibnamefont {Kasahara}}, \bibinfo {author}
  {\bibfnamefont {T.}~\bibnamefont {Watashige}}, \bibinfo {author}
  {\bibfnamefont {T.}~\bibnamefont {Shibauchi}}, \bibinfo {author}
  {\bibfnamefont {Y.}~\bibnamefont {Matsuda}}, \ and\ \bibinfo {author}
  {\bibfnamefont {O.~M.}\ \bibnamefont {Auslaender}},\ }\href {\doibase
  10.1103/PhysRevB.94.064510} {\bibfield  {journal} {\bibinfo  {journal} {Phys.
  Rev. B}\ }\textbf {\bibinfo {volume} {94}},\ \bibinfo {pages} {064510}
  (\bibinfo {year} {2016})}\BibitemShut {NoStop}%
\bibitem [{\citenamefont {Pratt}\ \emph {et~al.}(2009)\citenamefont {Pratt},
  \citenamefont {Tian}, \citenamefont {Kreyssig}, \citenamefont {Zarestky},
  \citenamefont {Nandi}, \citenamefont {Ni}, \citenamefont {Bud’ko},
  \citenamefont {Canfield}, \citenamefont {Goldman},\ and\ \citenamefont
  {McQueeney}}]{Pratt2009}%
  \BibitemOpen
  \bibfield  {author} {\bibinfo {author} {\bibfnamefont {D.~K.}\ \bibnamefont
  {Pratt}}, \bibinfo {author} {\bibfnamefont {W.}~\bibnamefont {Tian}},
  \bibinfo {author} {\bibfnamefont {A.}~\bibnamefont {Kreyssig}}, \bibinfo
  {author} {\bibfnamefont {J.~L.}\ \bibnamefont {Zarestky}}, \bibinfo {author}
  {\bibfnamefont {S.}~\bibnamefont {Nandi}}, \bibinfo {author} {\bibfnamefont
  {N.}~\bibnamefont {Ni}}, \bibinfo {author} {\bibfnamefont {S.~L.}\
  \bibnamefont {Bud’ko}}, \bibinfo {author} {\bibfnamefont {P.~C.}\
  \bibnamefont {Canfield}}, \bibinfo {author} {\bibfnamefont {A.~I.}\
  \bibnamefont {Goldman}}, \ and\ \bibinfo {author} {\bibfnamefont {R.~J.}\
  \bibnamefont {McQueeney}},\ }\href {\doibase 10.1103/PhysRevLett.103.087001}
  {\bibfield  {journal} {\bibinfo  {journal} {Phys. Rev. Lett.}\ }\textbf
  {\bibinfo {volume} {103}},\ \bibinfo {pages} {087001} (\bibinfo {year}
  {2009})}\BibitemShut {NoStop}%
\bibitem [{\citenamefont {Nandi}\ \emph {et~al.}(2010)\citenamefont {Nandi},
  \citenamefont {Kim}, \citenamefont {Kreyssig}, \citenamefont {Fernandes},
  \citenamefont {Pratt}, \citenamefont {Thaler}, \citenamefont {Ni},
  \citenamefont {Bud’ko}, \citenamefont {Canfield}, \citenamefont
  {Schmalian}, \citenamefont {McQueeney},\ and\ \citenamefont
  {Goldman}}]{Nandi2010}%
  \BibitemOpen
  \bibfield  {author} {\bibinfo {author} {\bibfnamefont {S.}~\bibnamefont
  {Nandi}}, \bibinfo {author} {\bibfnamefont {M.~G.}\ \bibnamefont {Kim}},
  \bibinfo {author} {\bibfnamefont {A.}~\bibnamefont {Kreyssig}}, \bibinfo
  {author} {\bibfnamefont {R.~M.}\ \bibnamefont {Fernandes}}, \bibinfo {author}
  {\bibfnamefont {D.~K.}\ \bibnamefont {Pratt}}, \bibinfo {author}
  {\bibfnamefont {A.}~\bibnamefont {Thaler}}, \bibinfo {author} {\bibfnamefont
  {N.}~\bibnamefont {Ni}}, \bibinfo {author} {\bibfnamefont {S.~L.}\
  \bibnamefont {Bud’ko}}, \bibinfo {author} {\bibfnamefont {P.~C.}\
  \bibnamefont {Canfield}}, \bibinfo {author} {\bibfnamefont {J.}~\bibnamefont
  {Schmalian}}, \bibinfo {author} {\bibfnamefont {R.~J.}\ \bibnamefont
  {McQueeney}}, \ and\ \bibinfo {author} {\bibfnamefont {A.~I.}\ \bibnamefont
  {Goldman}},\ }\href {\doibase 10.1103/PhysRevLett.104.057006} {\bibfield
  {journal} {\bibinfo  {journal} {Phys. Rev. Lett.}\ }\textbf {\bibinfo
  {volume} {104}},\ \bibinfo {pages} {057006} (\bibinfo {year}
  {2010})}\BibitemShut {NoStop}%
\bibitem [{\citenamefont {Kasahara}\ \emph {et~al.}(2012)\citenamefont
  {Kasahara}, \citenamefont {Shi}, \citenamefont {Hashimoto}, \citenamefont
  {Tonegawa}, \citenamefont {Mizukami}, \citenamefont {Shibauchi},
  \citenamefont {Sugimoto}, \citenamefont {Fukuda}, \citenamefont {Terashima},
  \citenamefont {Nevidomskyy},\ and\ \citenamefont {Matsuda}}]{Kasahara2012}%
  \BibitemOpen
  \bibfield  {author} {\bibinfo {author} {\bibfnamefont {S.}~\bibnamefont
  {Kasahara}}, \bibinfo {author} {\bibfnamefont {H.~J.}\ \bibnamefont {Shi}},
  \bibinfo {author} {\bibfnamefont {K.}~\bibnamefont {Hashimoto}}, \bibinfo
  {author} {\bibfnamefont {S.}~\bibnamefont {Tonegawa}}, \bibinfo {author}
  {\bibfnamefont {Y.}~\bibnamefont {Mizukami}}, \bibinfo {author}
  {\bibfnamefont {T.}~\bibnamefont {Shibauchi}}, \bibinfo {author}
  {\bibfnamefont {K.}~\bibnamefont {Sugimoto}}, \bibinfo {author}
  {\bibfnamefont {T.}~\bibnamefont {Fukuda}}, \bibinfo {author} {\bibfnamefont
  {T.}~\bibnamefont {Terashima}}, \bibinfo {author} {\bibfnamefont {A.~H.}\
  \bibnamefont {Nevidomskyy}}, \ and\ \bibinfo {author} {\bibfnamefont
  {Y.}~\bibnamefont {Matsuda}},\ }\href {\doibase 10.1038/nature11178}
  {\bibfield  {journal} {\bibinfo  {journal} {Nature}\ }\textbf {\bibinfo
  {volume} {486}},\ \bibinfo {pages} {382} (\bibinfo {year}
  {2012})}\BibitemShut {NoStop}%
\bibitem [{\citenamefont {B\"ohmer}\ \emph {et~al.}(2012)\citenamefont
  {B\"ohmer}, \citenamefont {Burger}, \citenamefont {Hardy}, \citenamefont
  {Wolf}, \citenamefont {Schweiss}, \citenamefont {Fromknecht}, \citenamefont
  {v.~L\"ohneysen}, \citenamefont {Meingast}, \citenamefont {Mak},
  \citenamefont {Lortz}, \citenamefont {Kasahara}, \citenamefont {Terashima},
  \citenamefont {Shibauchi},\ and\ \citenamefont {Matsuda}}]{Bohemer2012}%
  \BibitemOpen
  \bibfield  {author} {\bibinfo {author} {\bibfnamefont {A.~E.}\ \bibnamefont
  {B\"ohmer}}, \bibinfo {author} {\bibfnamefont {P.}~\bibnamefont {Burger}},
  \bibinfo {author} {\bibfnamefont {F.}~\bibnamefont {Hardy}}, \bibinfo
  {author} {\bibfnamefont {T.}~\bibnamefont {Wolf}}, \bibinfo {author}
  {\bibfnamefont {P.}~\bibnamefont {Schweiss}}, \bibinfo {author}
  {\bibfnamefont {R.}~\bibnamefont {Fromknecht}}, \bibinfo {author}
  {\bibfnamefont {H.}~\bibnamefont {v.~L\"ohneysen}}, \bibinfo {author}
  {\bibfnamefont {C.}~\bibnamefont {Meingast}}, \bibinfo {author}
  {\bibfnamefont {H.~K.}\ \bibnamefont {Mak}}, \bibinfo {author} {\bibfnamefont
  {R.}~\bibnamefont {Lortz}}, \bibinfo {author} {\bibfnamefont
  {S.}~\bibnamefont {Kasahara}}, \bibinfo {author} {\bibfnamefont
  {T.}~\bibnamefont {Terashima}}, \bibinfo {author} {\bibfnamefont
  {T.}~\bibnamefont {Shibauchi}}, \ and\ \bibinfo {author} {\bibfnamefont
  {Y.}~\bibnamefont {Matsuda}},\ }\href {\doibase 10.1103/PhysRevB.86.094521}
  {\bibfield  {journal} {\bibinfo  {journal} {Phys. Rev. B}\ }\textbf {\bibinfo
  {volume} {86}},\ \bibinfo {pages} {094521} (\bibinfo {year}
  {2012})}\BibitemShut {NoStop}%
\bibitem [{\citenamefont {Iye}\ \emph {et~al.}(2012{\natexlab{a}})\citenamefont
  {Iye}, \citenamefont {Nakai}, \citenamefont {Kitagawa}, \citenamefont
  {Ishida}, \citenamefont {Kasahara}, \citenamefont {Shibauchi}, \citenamefont
  {Matsuda},\ and\ \citenamefont {Terashima}}]{Iye2012a}%
  \BibitemOpen
  \bibfield  {author} {\bibinfo {author} {\bibfnamefont {T.}~\bibnamefont
  {Iye}}, \bibinfo {author} {\bibfnamefont {Y.}~\bibnamefont {Nakai}}, \bibinfo
  {author} {\bibfnamefont {S.}~\bibnamefont {Kitagawa}}, \bibinfo {author}
  {\bibfnamefont {K.}~\bibnamefont {Ishida}}, \bibinfo {author} {\bibfnamefont
  {S.}~\bibnamefont {Kasahara}}, \bibinfo {author} {\bibfnamefont
  {T.}~\bibnamefont {Shibauchi}}, \bibinfo {author} {\bibfnamefont
  {Y.}~\bibnamefont {Matsuda}}, \ and\ \bibinfo {author} {\bibfnamefont
  {T.}~\bibnamefont {Terashima}},\ }\href {\doibase 10.1103/PhysRevB.85.184505}
  {\bibfield  {journal} {\bibinfo  {journal} {Phys. Rev. B}\ }\textbf {\bibinfo
  {volume} {85}},\ \bibinfo {pages} {184505} (\bibinfo {year}
  {2012}{\natexlab{a}})}\BibitemShut {NoStop}%
\bibitem [{\citenamefont {Iye}\ \emph {et~al.}(2012{\natexlab{b}})\citenamefont
  {Iye}, \citenamefont {Nakai}, \citenamefont {Kitagawa}, \citenamefont
  {Ishida}, \citenamefont {Kasahara}, \citenamefont {Shibauchi}, \citenamefont
  {Matsuda},\ and\ \citenamefont {Terashima}}]{Iye2012b}%
  \BibitemOpen
  \bibfield  {author} {\bibinfo {author} {\bibfnamefont {T.}~\bibnamefont
  {Iye}}, \bibinfo {author} {\bibfnamefont {Y.}~\bibnamefont {Nakai}}, \bibinfo
  {author} {\bibfnamefont {S.}~\bibnamefont {Kitagawa}}, \bibinfo {author}
  {\bibfnamefont {K.}~\bibnamefont {Ishida}}, \bibinfo {author} {\bibfnamefont
  {S.}~\bibnamefont {Kasahara}}, \bibinfo {author} {\bibfnamefont
  {T.}~\bibnamefont {Shibauchi}}, \bibinfo {author} {\bibfnamefont
  {Y.}~\bibnamefont {Matsuda}}, \ and\ \bibinfo {author} {\bibfnamefont
  {T.}~\bibnamefont {Terashima}},\ }\href {\doibase 10.1143/JPSJ.81.033701}
  {\bibfield  {journal} {\bibinfo  {journal} {J. Phys. Soc. Jpn.}\ }\textbf
  {\bibinfo {volume} {81}},\ \bibinfo {pages} {033701} (\bibinfo {year}
  {2012}{\natexlab{b}})}\BibitemShut {NoStop}%
\bibitem [{\citenamefont {Hu}\ \emph {et~al.}(2015)\citenamefont {Hu},
  \citenamefont {Lu}, \citenamefont {Zhang}, \citenamefont {Luo}, \citenamefont
  {Li}, \citenamefont {Wang}, \citenamefont {Chen}, \citenamefont {Han},
  \citenamefont {Banjara}, \citenamefont {Sapkota}, \citenamefont {Kreyssig},
  \citenamefont {Goldman}, \citenamefont {Yamani}, \citenamefont {Niedermayer},
  \citenamefont {Skoulatos}, \citenamefont {Georgii}, \citenamefont {Keller},
  \citenamefont {Wang}, \citenamefont {Yu},\ and\ \citenamefont
  {Dai}}]{Hu2015}%
  \BibitemOpen
  \bibfield  {author} {\bibinfo {author} {\bibfnamefont {D.}~\bibnamefont
  {Hu}}, \bibinfo {author} {\bibfnamefont {X.}~\bibnamefont {Lu}}, \bibinfo
  {author} {\bibfnamefont {W.}~\bibnamefont {Zhang}}, \bibinfo {author}
  {\bibfnamefont {H.}~\bibnamefont {Luo}}, \bibinfo {author} {\bibfnamefont
  {S.}~\bibnamefont {Li}}, \bibinfo {author} {\bibfnamefont {P.}~\bibnamefont
  {Wang}}, \bibinfo {author} {\bibfnamefont {G.}~\bibnamefont {Chen}}, \bibinfo
  {author} {\bibfnamefont {F.}~\bibnamefont {Han}}, \bibinfo {author}
  {\bibfnamefont {S.~R.}\ \bibnamefont {Banjara}}, \bibinfo {author}
  {\bibfnamefont {A.}~\bibnamefont {Sapkota}}, \bibinfo {author} {\bibfnamefont
  {A.}~\bibnamefont {Kreyssig}}, \bibinfo {author} {\bibfnamefont {A.~I.}\
  \bibnamefont {Goldman}}, \bibinfo {author} {\bibfnamefont {Z.}~\bibnamefont
  {Yamani}}, \bibinfo {author} {\bibfnamefont {C.}~\bibnamefont {Niedermayer}},
  \bibinfo {author} {\bibfnamefont {M.}~\bibnamefont {Skoulatos}}, \bibinfo
  {author} {\bibfnamefont {R.}~\bibnamefont {Georgii}}, \bibinfo {author}
  {\bibfnamefont {T.}~\bibnamefont {Keller}}, \bibinfo {author} {\bibfnamefont
  {P.}~\bibnamefont {Wang}}, \bibinfo {author} {\bibfnamefont {W.}~\bibnamefont
  {Yu}}, \ and\ \bibinfo {author} {\bibfnamefont {P.}~\bibnamefont {Dai}},\
  }\href {\doibase 10.1103/PhysRevLett.114.157002} {\bibfield  {journal}
  {\bibinfo  {journal} {Phys. Rev. Lett.}\ }\textbf {\bibinfo {volume} {114}},\
  \bibinfo {pages} {157002} (\bibinfo {year} {2015})}\BibitemShut {NoStop}%
\bibitem [{\citenamefont {Wiesenmayer}\ \emph {et~al.}(2011)\citenamefont
  {Wiesenmayer}, \citenamefont {Luetkens}, \citenamefont {Pascua},
  \citenamefont {Khasanov}, \citenamefont {Amato}, \citenamefont {Potts},
  \citenamefont {Banusch}, \citenamefont {{Hans-Henning Klauss}},\ and\
  \citenamefont {Johrendt}}]{Wiesenmayer2011}%
  \BibitemOpen
  \bibfield  {author} {\bibinfo {author} {\bibfnamefont {E.}~\bibnamefont
  {Wiesenmayer}}, \bibinfo {author} {\bibfnamefont {H.}~\bibnamefont
  {Luetkens}}, \bibinfo {author} {\bibfnamefont {G.}~\bibnamefont {Pascua}},
  \bibinfo {author} {\bibfnamefont {R.}~\bibnamefont {Khasanov}}, \bibinfo
  {author} {\bibfnamefont {A.}~\bibnamefont {Amato}}, \bibinfo {author}
  {\bibfnamefont {H.}~\bibnamefont {Potts}}, \bibinfo {author} {\bibfnamefont
  {B.}~\bibnamefont {Banusch}}, \bibinfo {author} {\bibnamefont {{Hans-Henning
  Klauss}}}, \ and\ \bibinfo {author} {\bibfnamefont {D.}~\bibnamefont
  {Johrendt}},\ }\href {\doibase 10.1103/PhysRevLett.107.237001} {\bibfield
  {journal} {\bibinfo  {journal} {Phys. Rev. Lett.}\ }\textbf {\bibinfo
  {volume} {107}},\ \bibinfo {pages} {237001} (\bibinfo {year}
  {2011})}\BibitemShut {NoStop}%
\bibitem [{\citenamefont {Mallett}\ \emph {et~al.}(2017)\citenamefont
  {Mallett}, \citenamefont {Wang}, \citenamefont {Marsik}, \citenamefont
  {Sheveleva}, \citenamefont {Yazdi-Rizi}, \citenamefont {Tallon},
  \citenamefont {Adelmann}, \citenamefont {Wolf},\ and\ \citenamefont
  {Bernhard}}]{Mallett2017}%
  \BibitemOpen
  \bibfield  {author} {\bibinfo {author} {\bibfnamefont {B.~P.~P.}\
  \bibnamefont {Mallett}}, \bibinfo {author} {\bibfnamefont {C.~N.}\
  \bibnamefont {Wang}}, \bibinfo {author} {\bibfnamefont {P.}~\bibnamefont
  {Marsik}}, \bibinfo {author} {\bibfnamefont {E.}~\bibnamefont {Sheveleva}},
  \bibinfo {author} {\bibfnamefont {M.}~\bibnamefont {Yazdi-Rizi}}, \bibinfo
  {author} {\bibfnamefont {J.~L.}\ \bibnamefont {Tallon}}, \bibinfo {author}
  {\bibfnamefont {P.}~\bibnamefont {Adelmann}}, \bibinfo {author}
  {\bibfnamefont {T.}~\bibnamefont {Wolf}}, \ and\ \bibinfo {author}
  {\bibfnamefont {C.}~\bibnamefont {Bernhard}},\ }\href
  {https://journals.aps.org/prb/abstract/10.1103/PhysRevB.95.054512} {\bibfield
   {journal} {\bibinfo  {journal} {Phys. Rev. B}\ }\textbf {\bibinfo {volume}
  {95}},\ \bibinfo {pages} {054512} (\bibinfo {year} {2017})}\BibitemShut
  {NoStop}%
\bibitem [{\citenamefont {Hardy}\ \emph {et~al.}(2016)\citenamefont {Hardy},
  \citenamefont {B{\"o}hmer}, \citenamefont {{de' Medici}}, \citenamefont
  {Capone}, \citenamefont {Giovannetti}, \citenamefont {Eder}, \citenamefont
  {Wang}, \citenamefont {He}, \citenamefont {Wolf}, \citenamefont {Schweiss}
  \emph {et~al.}}]{Hardy2016}%
  \BibitemOpen
  \bibfield  {author} {\bibinfo {author} {\bibfnamefont {F.}~\bibnamefont
  {Hardy}}, \bibinfo {author} {\bibfnamefont {A.~E.}\ \bibnamefont
  {B{\"o}hmer}}, \bibinfo {author} {\bibfnamefont {L.}~\bibnamefont {{de'
  Medici}}}, \bibinfo {author} {\bibfnamefont {M.}~\bibnamefont {Capone}},
  \bibinfo {author} {\bibfnamefont {G.}~\bibnamefont {Giovannetti}}, \bibinfo
  {author} {\bibfnamefont {R.}~\bibnamefont {Eder}}, \bibinfo {author}
  {\bibfnamefont {L.}~\bibnamefont {Wang}}, \bibinfo {author} {\bibfnamefont
  {M.}~\bibnamefont {He}}, \bibinfo {author} {\bibfnamefont {T.}~\bibnamefont
  {Wolf}}, \bibinfo {author} {\bibfnamefont {P.}~\bibnamefont {Schweiss}},
  \emph {et~al.},\ }\href {http://link.aps.org/doi/10.1103/PhysRevB.94.205113}
  {\bibfield  {journal} {\bibinfo  {journal} {Phys. Rev. B}\ }\textbf {\bibinfo
  {volume} {94}},\ \bibinfo {pages} {205113} (\bibinfo {year}
  {2016})}\BibitemShut {NoStop}%
\bibitem [{\citenamefont {Cho}\ \emph {et~al.}(2016)\citenamefont {Cho},
  \citenamefont {Konczykowski}, \citenamefont {Teknowijoyo}, \citenamefont
  {Tanatar}, \citenamefont {Liu}, \citenamefont {Lograsso}, \citenamefont
  {Straszheim}, \citenamefont {Mishra}, \citenamefont {Maiti}, \citenamefont
  {Hirschfeld},\ and\ \citenamefont {Prozorov}}]{Cho2016}%
  \BibitemOpen
  \bibfield  {author} {\bibinfo {author} {\bibfnamefont {K.}~\bibnamefont
  {Cho}}, \bibinfo {author} {\bibfnamefont {M.}~\bibnamefont {Konczykowski}},
  \bibinfo {author} {\bibfnamefont {S.}~\bibnamefont {Teknowijoyo}}, \bibinfo
  {author} {\bibfnamefont {M.~A.}\ \bibnamefont {Tanatar}}, \bibinfo {author}
  {\bibfnamefont {Y.}~\bibnamefont {Liu}}, \bibinfo {author} {\bibfnamefont
  {T.~A.}\ \bibnamefont {Lograsso}}, \bibinfo {author} {\bibfnamefont {W.~E.}\
  \bibnamefont {Straszheim}}, \bibinfo {author} {\bibfnamefont
  {V.}~\bibnamefont {Mishra}}, \bibinfo {author} {\bibfnamefont
  {S.}~\bibnamefont {Maiti}}, \bibinfo {author} {\bibfnamefont {P.~J.}\
  \bibnamefont {Hirschfeld}}, \ and\ \bibinfo {author} {\bibfnamefont
  {R.}~\bibnamefont {Prozorov}},\ }\href
  {http://advances.sciencemag.org/content/2/9/e1600807} {\bibfield  {journal}
  {\bibinfo  {journal} {Sci. Adv.}\ }\textbf {\bibinfo {volume} {2}},\ \bibinfo
  {pages} {e1600807} (\bibinfo {year} {2016})}\BibitemShut {NoStop}%
\bibitem [{\citenamefont {Mallett}\ \emph {et~al.}(2015)\citenamefont
  {Mallett}, \citenamefont {Marsik}, \citenamefont {Yazdi-Rizi}, \citenamefont
  {Wolf}, \citenamefont {B\"ohmer}, \citenamefont {Hardy}, \citenamefont
  {Meingast}, \citenamefont {Munzar},\ and\ \citenamefont
  {Bernhard}}]{Mallett2015}%
  \BibitemOpen
  \bibfield  {author} {\bibinfo {author} {\bibfnamefont {B.~P.~P.}\
  \bibnamefont {Mallett}}, \bibinfo {author} {\bibfnamefont {P.}~\bibnamefont
  {Marsik}}, \bibinfo {author} {\bibfnamefont {M.}~\bibnamefont {Yazdi-Rizi}},
  \bibinfo {author} {\bibfnamefont {T.}~\bibnamefont {Wolf}}, \bibinfo {author}
  {\bibfnamefont {A.~E.}\ \bibnamefont {B\"ohmer}}, \bibinfo {author}
  {\bibfnamefont {F.}~\bibnamefont {Hardy}}, \bibinfo {author} {\bibfnamefont
  {C.}~\bibnamefont {Meingast}}, \bibinfo {author} {\bibfnamefont
  {D.}~\bibnamefont {Munzar}}, \ and\ \bibinfo {author} {\bibfnamefont
  {C.}~\bibnamefont {Bernhard}},\ }\href {\doibase
  10.1103/PhysRevLett.115.027003} {\bibfield  {journal} {\bibinfo  {journal}
  {Phys. Rev. Lett.}\ }\textbf {\bibinfo {volume} {115}},\ \bibinfo {pages}
  {027003} (\bibinfo {year} {2015})}\BibitemShut {NoStop}%
\bibitem [{\citenamefont {Malaeb}\ \emph {et~al.}(2012)\citenamefont {Malaeb},
  \citenamefont {Shimojima}, \citenamefont {Ishida}, \citenamefont {Okazaki},
  \citenamefont {Ota}, \citenamefont {Ohgushi}, \citenamefont {Kihou},
  \citenamefont {Saito}, \citenamefont {Lee}, \citenamefont {Ishida},
  \citenamefont {Nakajima}, \citenamefont {Uchida}, \citenamefont {Fukazawa},
  \citenamefont {Kohori}, \citenamefont {Iyo}, \citenamefont {Eisaki},
  \citenamefont {Chen}, \citenamefont {Watanabe}, \citenamefont {Ikeda},\ and\
  \citenamefont {Shin}}]{Malaeb2012}%
  \BibitemOpen
  \bibfield  {author} {\bibinfo {author} {\bibfnamefont {W.}~\bibnamefont
  {Malaeb}}, \bibinfo {author} {\bibfnamefont {T.}~\bibnamefont {Shimojima}},
  \bibinfo {author} {\bibfnamefont {Y.}~\bibnamefont {Ishida}}, \bibinfo
  {author} {\bibfnamefont {K.}~\bibnamefont {Okazaki}}, \bibinfo {author}
  {\bibfnamefont {Y.}~\bibnamefont {Ota}}, \bibinfo {author} {\bibfnamefont
  {K.}~\bibnamefont {Ohgushi}}, \bibinfo {author} {\bibfnamefont
  {K.}~\bibnamefont {Kihou}}, \bibinfo {author} {\bibfnamefont
  {T.}~\bibnamefont {Saito}}, \bibinfo {author} {\bibfnamefont {C.~H.}\
  \bibnamefont {Lee}}, \bibinfo {author} {\bibfnamefont {S.}~\bibnamefont
  {Ishida}}, \bibinfo {author} {\bibfnamefont {M.}~\bibnamefont {Nakajima}},
  \bibinfo {author} {\bibfnamefont {S.}~\bibnamefont {Uchida}}, \bibinfo
  {author} {\bibfnamefont {H.}~\bibnamefont {Fukazawa}}, \bibinfo {author}
  {\bibfnamefont {Y.}~\bibnamefont {Kohori}}, \bibinfo {author} {\bibfnamefont
  {A.}~\bibnamefont {Iyo}}, \bibinfo {author} {\bibfnamefont {H.}~\bibnamefont
  {Eisaki}}, \bibinfo {author} {\bibfnamefont {C.-T.}\ \bibnamefont {Chen}},
  \bibinfo {author} {\bibfnamefont {S.}~\bibnamefont {Watanabe}}, \bibinfo
  {author} {\bibfnamefont {H.}~\bibnamefont {Ikeda}}, \ and\ \bibinfo {author}
  {\bibfnamefont {S.}~\bibnamefont {Shin}},\ }\href {\doibase
  10.1103/PhysRevB.86.165117} {\bibfield  {journal} {\bibinfo  {journal} {Phys.
  Rev. B}\ }\textbf {\bibinfo {volume} {86}},\ \bibinfo {pages} {165117}
  (\bibinfo {year} {2012})}\BibitemShut {NoStop}%
\bibitem [{\citenamefont {Goko}\ \emph {et~al.}(2009)\citenamefont {Goko},
  \citenamefont {Aczel}, \citenamefont {Baggio-Saitovitch}, \citenamefont
  {Bud'ko}, \citenamefont {Canfield}, \citenamefont {Carlo}, \citenamefont
  {Chen}, \citenamefont {Dai}, \citenamefont {Hamann}, \citenamefont {Hu},
  \citenamefont {Kageyama}, \citenamefont {Luke}, \citenamefont {Luo},
  \citenamefont {Nachumi}, \citenamefont {Ni}, \citenamefont {Reznik},
  \citenamefont {Sanchez-Candela}, \citenamefont {Savici}, \citenamefont
  {Sikes}, \citenamefont {Wang}, \citenamefont {Wiebe}, \citenamefont
  {Williams}, \citenamefont {Yamamoto}, \citenamefont {Yu},\ and\ \citenamefont
  {Uemura}}]{Goko2009}%
  \BibitemOpen
  \bibfield  {author} {\bibinfo {author} {\bibfnamefont {T.}~\bibnamefont
  {Goko}}, \bibinfo {author} {\bibfnamefont {A.~A.}\ \bibnamefont {Aczel}},
  \bibinfo {author} {\bibfnamefont {E.}~\bibnamefont {Baggio-Saitovitch}},
  \bibinfo {author} {\bibfnamefont {S.~L.}\ \bibnamefont {Bud'ko}}, \bibinfo
  {author} {\bibfnamefont {P.~C.}\ \bibnamefont {Canfield}}, \bibinfo {author}
  {\bibfnamefont {J.~P.}\ \bibnamefont {Carlo}}, \bibinfo {author}
  {\bibfnamefont {G.~F.}\ \bibnamefont {Chen}}, \bibinfo {author}
  {\bibfnamefont {P.}~\bibnamefont {Dai}}, \bibinfo {author} {\bibfnamefont
  {A.~C.}\ \bibnamefont {Hamann}}, \bibinfo {author} {\bibfnamefont {W.~Z.}\
  \bibnamefont {Hu}}, \bibinfo {author} {\bibfnamefont {H.}~\bibnamefont
  {Kageyama}}, \bibinfo {author} {\bibfnamefont {G.~M.}\ \bibnamefont {Luke}},
  \bibinfo {author} {\bibfnamefont {J.~L.}\ \bibnamefont {Luo}}, \bibinfo
  {author} {\bibfnamefont {B.}~\bibnamefont {Nachumi}}, \bibinfo {author}
  {\bibfnamefont {N.}~\bibnamefont {Ni}}, \bibinfo {author} {\bibfnamefont
  {D.}~\bibnamefont {Reznik}}, \bibinfo {author} {\bibfnamefont {D.~R.}\
  \bibnamefont {Sanchez-Candela}}, \bibinfo {author} {\bibfnamefont {A.~T.}\
  \bibnamefont {Savici}}, \bibinfo {author} {\bibfnamefont {K.~J.}\
  \bibnamefont {Sikes}}, \bibinfo {author} {\bibfnamefont {N.~L.}\ \bibnamefont
  {Wang}}, \bibinfo {author} {\bibfnamefont {C.~R.}\ \bibnamefont {Wiebe}},
  \bibinfo {author} {\bibfnamefont {T.~J.}\ \bibnamefont {Williams}}, \bibinfo
  {author} {\bibfnamefont {T.}~\bibnamefont {Yamamoto}}, \bibinfo {author}
  {\bibfnamefont {W.}~\bibnamefont {Yu}}, \ and\ \bibinfo {author}
  {\bibfnamefont {Y.~J.}\ \bibnamefont {Uemura}},\ }\href {\doibase
  10.1103/PhysRevB.80.024508} {\bibfield  {journal} {\bibinfo  {journal} {Phys.
  Rev. B}\ }\textbf {\bibinfo {volume} {80}},\ \bibinfo {pages} {024508}
  (\bibinfo {year} {2009})}\BibitemShut {NoStop}%
\bibitem [{\citenamefont {Mu}\ \emph {et~al.}(2009)\citenamefont {Mu},
  \citenamefont {Luo}, \citenamefont {Wang}, \citenamefont {Shan},
  \citenamefont {Ren},\ and\ \citenamefont {Wen}}]{Mu2009}%
  \BibitemOpen
  \bibfield  {author} {\bibinfo {author} {\bibfnamefont {G.}~\bibnamefont
  {Mu}}, \bibinfo {author} {\bibfnamefont {H.}~\bibnamefont {Luo}}, \bibinfo
  {author} {\bibfnamefont {Z.}~\bibnamefont {Wang}}, \bibinfo {author}
  {\bibfnamefont {L.}~\bibnamefont {Shan}}, \bibinfo {author} {\bibfnamefont
  {C.}~\bibnamefont {Ren}}, \ and\ \bibinfo {author} {\bibfnamefont {H.~H.}\
  \bibnamefont {Wen}},\ }\href {\doibase 10.1103/PhysRevB.79.174501} {\bibfield
   {journal} {\bibinfo  {journal} {Phys. Rev. B}\ }\textbf {\bibinfo {volume}
  {79}},\ \bibinfo {pages} {174501} (\bibinfo {year} {2009})}\BibitemShut
  {NoStop}%
\bibitem [{\citenamefont {Xu}\ \emph {et~al.}(2013)\citenamefont {Xu},
  \citenamefont {Richard}, \citenamefont {Shi}, \citenamefont {van Roekeghem},
  \citenamefont {Qian}, \citenamefont {Razzoli}, \citenamefont {Rienks},
  \citenamefont {Chen}, \citenamefont {Ieki}, \citenamefont {Nakayama} \emph
  {et~al.}}]{Xu2013}%
  \BibitemOpen
  \bibfield  {author} {\bibinfo {author} {\bibfnamefont {N.}~\bibnamefont
  {Xu}}, \bibinfo {author} {\bibfnamefont {P.}~\bibnamefont {Richard}},
  \bibinfo {author} {\bibfnamefont {X.}~\bibnamefont {Shi}}, \bibinfo {author}
  {\bibfnamefont {A.}~\bibnamefont {van Roekeghem}}, \bibinfo {author}
  {\bibfnamefont {T.}~\bibnamefont {Qian}}, \bibinfo {author} {\bibfnamefont
  {E.}~\bibnamefont {Razzoli}}, \bibinfo {author} {\bibfnamefont
  {E.}~\bibnamefont {Rienks}}, \bibinfo {author} {\bibfnamefont {G.~F.}\
  \bibnamefont {Chen}}, \bibinfo {author} {\bibfnamefont {E.}~\bibnamefont
  {Ieki}}, \bibinfo {author} {\bibfnamefont {K.}~\bibnamefont {Nakayama}},
  \emph {et~al.},\ }\href {http://link.aps.org/doi/10.1103/PhysRevB.88.220508}
  {\bibfield  {journal} {\bibinfo  {journal} {Phys. Rev. B}\ }\textbf {\bibinfo
  {volume} {88}},\ \bibinfo {pages} {220508} (\bibinfo {year}
  {2013})}\BibitemShut {NoStop}%
\bibitem [{\citenamefont {Sefat}\ \emph {et~al.}(2008)\citenamefont {Sefat},
  \citenamefont {Jin}, \citenamefont {McGuire}, \citenamefont {Sales},
  \citenamefont {Singh},\ and\ \citenamefont {Mandrus}}]{Athena2008}%
  \BibitemOpen
  \bibfield  {author} {\bibinfo {author} {\bibfnamefont {A.~S.}\ \bibnamefont
  {Sefat}}, \bibinfo {author} {\bibfnamefont {R.}~\bibnamefont {Jin}}, \bibinfo
  {author} {\bibfnamefont {M.~A.}\ \bibnamefont {McGuire}}, \bibinfo {author}
  {\bibfnamefont {B.~C.}\ \bibnamefont {Sales}}, \bibinfo {author}
  {\bibfnamefont {D.~J.}\ \bibnamefont {Singh}}, \ and\ \bibinfo {author}
  {\bibfnamefont {D.}~\bibnamefont {Mandrus}},\ }\href {\doibase
  10.1103/PhysRevLett.101.117004} {\bibfield  {journal} {\bibinfo  {journal}
  {Phys. Rev. Lett.}\ }\textbf {\bibinfo {volume} {101}},\ \bibinfo {pages}
  {117004} (\bibinfo {year} {2008})}\BibitemShut {NoStop}%
\bibitem [{\citenamefont {Rotter}\ \emph {et~al.}(2008)\citenamefont {Rotter},
  \citenamefont {Tegel},\ and\ \citenamefont {Johrendt}}]{Rotter2008}%
  \BibitemOpen
  \bibfield  {author} {\bibinfo {author} {\bibfnamefont {M.}~\bibnamefont
  {Rotter}}, \bibinfo {author} {\bibfnamefont {M.}~\bibnamefont {Tegel}}, \
  and\ \bibinfo {author} {\bibfnamefont {D.}~\bibnamefont {Johrendt}},\ }\href
  {\doibase 10.1103/PhysRevLett.101.107006} {\bibfield  {journal} {\bibinfo
  {journal} {Phys. Rev. Lett.}\ }\textbf {\bibinfo {volume} {101}},\ \bibinfo
  {pages} {107006} (\bibinfo {year} {2008})}\BibitemShut {NoStop}%
\bibitem [{\citenamefont {Rotter}\ \emph {et~al.}(2010)\citenamefont {Rotter},
  \citenamefont {Hieke},\ and\ \citenamefont {Johrendt}}]{Rotter2010}%
  \BibitemOpen
  \bibfield  {author} {\bibinfo {author} {\bibfnamefont {M.}~\bibnamefont
  {Rotter}}, \bibinfo {author} {\bibfnamefont {C.}~\bibnamefont {Hieke}}, \
  and\ \bibinfo {author} {\bibfnamefont {D.}~\bibnamefont {Johrendt}},\ }\href
  {\doibase 10.1103/PhysRevB.82.014513} {\bibfield  {journal} {\bibinfo
  {journal} {Phys. Rev. B}\ }\textbf {\bibinfo {volume} {82}},\ \bibinfo
  {pages} {014513} (\bibinfo {year} {2010})}\BibitemShut {NoStop}%
\bibitem [{\citenamefont {Khasanov}\ \emph {et~al.}(2009)\citenamefont
  {Khasanov}, \citenamefont {Evtushinsky}, \citenamefont {Amato}, \citenamefont
  {Klauss}, \citenamefont {Luetkens}, \citenamefont {Niedermayer},
  \citenamefont {B\"{u}chner}, \citenamefont {Sun}, \citenamefont {Lin},
  \citenamefont {Park}, \citenamefont {Inosov},\ and\ \citenamefont
  {Hinkov}}]{Khasanov2009}%
  \BibitemOpen
  \bibfield  {author} {\bibinfo {author} {\bibfnamefont {R.}~\bibnamefont
  {Khasanov}}, \bibinfo {author} {\bibfnamefont {D.~V.}\ \bibnamefont
  {Evtushinsky}}, \bibinfo {author} {\bibfnamefont {A.}~\bibnamefont {Amato}},
  \bibinfo {author} {\bibfnamefont {H.~H.}\ \bibnamefont {Klauss}}, \bibinfo
  {author} {\bibfnamefont {H.}~\bibnamefont {Luetkens}}, \bibinfo {author}
  {\bibfnamefont {C.}~\bibnamefont {Niedermayer}}, \bibinfo {author}
  {\bibfnamefont {B.}~\bibnamefont {B\"{u}chner}}, \bibinfo {author}
  {\bibfnamefont {G.~L.}\ \bibnamefont {Sun}}, \bibinfo {author} {\bibfnamefont
  {C.~T.}\ \bibnamefont {Lin}}, \bibinfo {author} {\bibfnamefont {J.~T.}\
  \bibnamefont {Park}}, \bibinfo {author} {\bibfnamefont {D.~S.}\ \bibnamefont
  {Inosov}}, \ and\ \bibinfo {author} {\bibfnamefont {V.}~\bibnamefont
  {Hinkov}},\ }\href {\doibase 10.1103/PhysRevLett.102.187005} {\bibfield
  {journal} {\bibinfo  {journal} {Phys. Rev. Lett.}\ }\textbf {\bibinfo
  {volume} {102}},\ \bibinfo {pages} {187005} (\bibinfo {year}
  {2009})}\BibitemShut {NoStop}%
\bibitem [{\citenamefont {Ohishi}\ \emph {et~al.}(2012)\citenamefont {Ohishi},
  \citenamefont {Ishii}, \citenamefont {Watanabe}, \citenamefont {Fukazawa},
  \citenamefont {Saito}, \citenamefont {Kohori}, \citenamefont {Kihou},
  \citenamefont {Lee}, \citenamefont {Kito}, \citenamefont {Iyo},\ and\
  \citenamefont {Eisaki}}]{Ohishi2012}%
  \BibitemOpen
  \bibfield  {author} {\bibinfo {author} {\bibfnamefont {K.}~\bibnamefont
  {Ohishi}}, \bibinfo {author} {\bibfnamefont {Y.}~\bibnamefont {Ishii}},
  \bibinfo {author} {\bibfnamefont {I.}~\bibnamefont {Watanabe}}, \bibinfo
  {author} {\bibfnamefont {H.}~\bibnamefont {Fukazawa}}, \bibinfo {author}
  {\bibfnamefont {T.}~\bibnamefont {Saito}}, \bibinfo {author} {\bibfnamefont
  {Y.}~\bibnamefont {Kohori}}, \bibinfo {author} {\bibfnamefont
  {K.}~\bibnamefont {Kihou}}, \bibinfo {author} {\bibfnamefont
  {C.}~\bibnamefont {Lee}}, \bibinfo {author} {\bibfnamefont {H.}~\bibnamefont
  {Kito}}, \bibinfo {author} {\bibfnamefont {A.}~\bibnamefont {Iyo}}, \ and\
  \bibinfo {author} {\bibfnamefont {H.}~\bibnamefont {Eisaki}},\ }\href
  {\doibase 10.1143/JPSJS.81SB.SB046} {\bibfield  {journal} {\bibinfo
  {journal} {J. Phys. Soc. Jpn.}\ }\textbf {\bibinfo {volume} {81}},\ \bibinfo
  {pages} {SB046} (\bibinfo {year} {2012})}\BibitemShut {NoStop}%
\bibitem [{\citenamefont {Cho}\ \emph {et~al.}(2014)\citenamefont {Cho},
  \citenamefont {Ko\'nczykowski}, \citenamefont {Murphy}, \citenamefont {Kim},
  \citenamefont {Tanatar}, \citenamefont {Straszheim}, \citenamefont {Shen},
  \citenamefont {Wen},\ and\ \citenamefont {Prozorov}}]{Cho2014}%
  \BibitemOpen
  \bibfield  {author} {\bibinfo {author} {\bibfnamefont {K.}~\bibnamefont
  {Cho}}, \bibinfo {author} {\bibfnamefont {M.}~\bibnamefont {Ko\'nczykowski}},
  \bibinfo {author} {\bibfnamefont {J.}~\bibnamefont {Murphy}}, \bibinfo
  {author} {\bibfnamefont {H.}~\bibnamefont {Kim}}, \bibinfo {author}
  {\bibfnamefont {M.~A.}\ \bibnamefont {Tanatar}}, \bibinfo {author}
  {\bibfnamefont {W.~E.}\ \bibnamefont {Straszheim}}, \bibinfo {author}
  {\bibfnamefont {B.}~\bibnamefont {Shen}}, \bibinfo {author} {\bibfnamefont
  {H.~H.}\ \bibnamefont {Wen}}, \ and\ \bibinfo {author} {\bibfnamefont
  {R.}~\bibnamefont {Prozorov}},\ }\href {\doibase 10.1103/PhysRevB.90.104514}
  {\bibfield  {journal} {\bibinfo  {journal} {Phys. Rev. B}\ }\textbf {\bibinfo
  {volume} {90}},\ \bibinfo {pages} {104514} (\bibinfo {year}
  {2014})}\BibitemShut {NoStop}%
\bibitem [{\citenamefont {Khan}\ and\ \citenamefont
  {Johnson}(2014)}]{Khan2014}%
  \BibitemOpen
  \bibfield  {author} {\bibinfo {author} {\bibfnamefont {S.~N.}\ \bibnamefont
  {Khan}}\ and\ \bibinfo {author} {\bibfnamefont {D.~D.}\ \bibnamefont
  {Johnson}},\ }\href {\doibase 10.1103/PhysRevLett.112.156401} {\bibfield
  {journal} {\bibinfo  {journal} {Phys. Rev. Lett.}\ }\textbf {\bibinfo
  {volume} {112}},\ \bibinfo {pages} {156401} (\bibinfo {year}
  {2014})}\BibitemShut {NoStop}%
\bibitem [{\citenamefont {Liu}\ \emph {et~al.}(2014)\citenamefont {Liu},
  \citenamefont {Tanatar}, \citenamefont {Straszheim}, \citenamefont {Jensen},
  \citenamefont {Dennis}, \citenamefont {McCallum}, \citenamefont {Kogan},
  \citenamefont {Prozorov},\ and\ \citenamefont {Lograsso}}]{Liu2014}%
  \BibitemOpen
  \bibfield  {author} {\bibinfo {author} {\bibfnamefont {Y.}~\bibnamefont
  {Liu}}, \bibinfo {author} {\bibfnamefont {M.~A.}\ \bibnamefont {Tanatar}},
  \bibinfo {author} {\bibfnamefont {W.~E.}\ \bibnamefont {Straszheim}},
  \bibinfo {author} {\bibfnamefont {B.}~\bibnamefont {Jensen}}, \bibinfo
  {author} {\bibfnamefont {K.~W.}\ \bibnamefont {Dennis}}, \bibinfo {author}
  {\bibfnamefont {R.~W.}\ \bibnamefont {McCallum}}, \bibinfo {author}
  {\bibfnamefont {V.~G.}\ \bibnamefont {Kogan}}, \bibinfo {author}
  {\bibfnamefont {R.}~\bibnamefont {Prozorov}}, \ and\ \bibinfo {author}
  {\bibfnamefont {T.~A.}\ \bibnamefont {Lograsso}},\ }\href {\doibase
  10.1103/PhysRevB.89.134504} {\bibfield  {journal} {\bibinfo  {journal} {Phys.
  Rev. B}\ }\textbf {\bibinfo {volume} {89}},\ \bibinfo {pages} {134504}
  (\bibinfo {year} {2014})}\BibitemShut {NoStop}%
\bibitem [{\citenamefont {Prozorov}\ and\ \citenamefont
  {Giannetta}(2006)}]{Prozorov2006}%
  \BibitemOpen
  \bibfield  {author} {\bibinfo {author} {\bibfnamefont {R.}~\bibnamefont
  {Prozorov}}\ and\ \bibinfo {author} {\bibfnamefont {R.~W.}\ \bibnamefont
  {Giannetta}},\ }\href {http://www.iop.org/EJ/abstract/0953-2048/19/8/R01/}
  {\bibfield  {journal} {\bibinfo  {journal} {Supercond. Sci. Technol.}\
  }\textbf {\bibinfo {volume} {19}},\ \bibinfo {pages} {R41} (\bibinfo {year}
  {2006})}\BibitemShut {NoStop}%
\bibitem [{\citenamefont {Martin}\ \emph {et~al.}(2009)\citenamefont {Martin},
  \citenamefont {Gordon}, \citenamefont {Tanatar}, \citenamefont {Kim},
  \citenamefont {Ni}, \citenamefont {Bud'ko}, \citenamefont {Canfield},
  \citenamefont {Luo}, \citenamefont {Wen}, \citenamefont {Wang}, \citenamefont
  {Vorontsov}, \citenamefont {Kogan},\ and\ \citenamefont
  {Prozorov}}]{Martin2009}%
  \BibitemOpen
  \bibfield  {author} {\bibinfo {author} {\bibfnamefont {C.}~\bibnamefont
  {Martin}}, \bibinfo {author} {\bibfnamefont {R.~T.}\ \bibnamefont {Gordon}},
  \bibinfo {author} {\bibfnamefont {M.~A.}\ \bibnamefont {Tanatar}}, \bibinfo
  {author} {\bibfnamefont {H.}~\bibnamefont {Kim}}, \bibinfo {author}
  {\bibfnamefont {N.}~\bibnamefont {Ni}}, \bibinfo {author} {\bibfnamefont
  {S.~L.}\ \bibnamefont {Bud'ko}}, \bibinfo {author} {\bibfnamefont {P.~C.}\
  \bibnamefont {Canfield}}, \bibinfo {author} {\bibfnamefont {H.}~\bibnamefont
  {Luo}}, \bibinfo {author} {\bibfnamefont {H.~H.}\ \bibnamefont {Wen}},
  \bibinfo {author} {\bibfnamefont {Z.}~\bibnamefont {Wang}}, \bibinfo {author}
  {\bibfnamefont {A.~B.}\ \bibnamefont {Vorontsov}}, \bibinfo {author}
  {\bibfnamefont {V.~G.}\ \bibnamefont {Kogan}}, \ and\ \bibinfo {author}
  {\bibfnamefont {R.}~\bibnamefont {Prozorov}},\ }\href
  {http://link.aps.org/doi/10.1103/PhysRevB.80.020501} {\bibfield  {journal}
  {\bibinfo  {journal} {Phys. Rev. B}\ }\textbf {\bibinfo {volume} {80}},\
  \bibinfo {pages} {020501} (\bibinfo {year} {2009})}\BibitemShut {NoStop}%
\bibitem [{\citenamefont {Kim}\ \emph {et~al.}(2014)\citenamefont {Kim},
  \citenamefont {Tanatar}, \citenamefont {Straszheim}, \citenamefont {Cho},
  \citenamefont {Murphy}, \citenamefont {Spyrison}, \citenamefont {Reid},
  \citenamefont {Shen}, \citenamefont {Wen}, \citenamefont {Fernandes},\ and\
  \citenamefont {Prozorov}}]{Kim2014}%
  \BibitemOpen
  \bibfield  {author} {\bibinfo {author} {\bibfnamefont {H.}~\bibnamefont
  {Kim}}, \bibinfo {author} {\bibfnamefont {M.~A.}\ \bibnamefont {Tanatar}},
  \bibinfo {author} {\bibfnamefont {W.~E.}\ \bibnamefont {Straszheim}},
  \bibinfo {author} {\bibfnamefont {K.}~\bibnamefont {Cho}}, \bibinfo {author}
  {\bibfnamefont {J.}~\bibnamefont {Murphy}}, \bibinfo {author} {\bibfnamefont
  {N.}~\bibnamefont {Spyrison}}, \bibinfo {author} {\bibfnamefont {J.-P.}\
  \bibnamefont {Reid}}, \bibinfo {author} {\bibfnamefont {B.}~\bibnamefont
  {Shen}}, \bibinfo {author} {\bibfnamefont {H.-H.}\ \bibnamefont {Wen}},
  \bibinfo {author} {\bibfnamefont {R.~M.}\ \bibnamefont {Fernandes}}, \ and\
  \bibinfo {author} {\bibfnamefont {R.}~\bibnamefont {Prozorov}},\ }\href
  {\doibase 10.1103/PhysRevB.90.014517} {\bibfield  {journal} {\bibinfo
  {journal} {Phys. Rev. B}\ }\textbf {\bibinfo {volume} {90}},\ \bibinfo
  {pages} {014517} (\bibinfo {year} {2014})}\BibitemShut {NoStop}%
\bibitem [{\citenamefont {Gordon}\ \emph
  {et~al.}(2010{\natexlab{a}})\citenamefont {Gordon}, \citenamefont {Kim},
  \citenamefont {Salovich}, \citenamefont {Giannetta}, \citenamefont
  {Fernandes}, \citenamefont {Kogan}, \citenamefont {Prozorov}, \citenamefont
  {Bud’ko}, \citenamefont {Canfield}, \citenamefont {Tanatar},\ and\
  \citenamefont {Prozorov}}]{Gordon2010}%
  \BibitemOpen
  \bibfield  {author} {\bibinfo {author} {\bibfnamefont {R.~T.}\ \bibnamefont
  {Gordon}}, \bibinfo {author} {\bibfnamefont {H.}~\bibnamefont {Kim}},
  \bibinfo {author} {\bibfnamefont {N.}~\bibnamefont {Salovich}}, \bibinfo
  {author} {\bibfnamefont {R.~W.}\ \bibnamefont {Giannetta}}, \bibinfo {author}
  {\bibfnamefont {R.~M.}\ \bibnamefont {Fernandes}}, \bibinfo {author}
  {\bibfnamefont {V.~G.}\ \bibnamefont {Kogan}}, \bibinfo {author}
  {\bibfnamefont {T.}~\bibnamefont {Prozorov}}, \bibinfo {author}
  {\bibfnamefont {S.~L.}\ \bibnamefont {Bud’ko}}, \bibinfo {author}
  {\bibfnamefont {P.~C.}\ \bibnamefont {Canfield}}, \bibinfo {author}
  {\bibfnamefont {M.~A.}\ \bibnamefont {Tanatar}}, \ and\ \bibinfo {author}
  {\bibfnamefont {R.}~\bibnamefont {Prozorov}},\ }\href {\doibase
  10.1103/PhysRevB.82.054507} {\bibfield  {journal} {\bibinfo  {journal} {Phys.
  Rev. B}\ }\textbf {\bibinfo {volume} {82}},\ \bibinfo {pages} {054507}
  (\bibinfo {year} {2010}{\natexlab{a}})}\BibitemShut {NoStop}%
\bibitem [{\citenamefont {Auslaender}\ \emph {et~al.}(2009)\citenamefont
  {Auslaender}, \citenamefont {Luan}, \citenamefont {Straver}, \citenamefont
  {Hoffman}, \citenamefont {Koshnick}, \citenamefont {Zeldov}, \citenamefont
  {Bonn}, \citenamefont {Liang}, \citenamefont {Hardy},\ and\ \citenamefont
  {Moler}}]{Auslaender2009}%
  \BibitemOpen
  \bibfield  {author} {\bibinfo {author} {\bibfnamefont {O.~M.}\ \bibnamefont
  {Auslaender}}, \bibinfo {author} {\bibfnamefont {L.}~\bibnamefont {Luan}},
  \bibinfo {author} {\bibfnamefont {E.~W.~J.}\ \bibnamefont {Straver}},
  \bibinfo {author} {\bibfnamefont {J.~E.}\ \bibnamefont {Hoffman}}, \bibinfo
  {author} {\bibfnamefont {N.~C.}\ \bibnamefont {Koshnick}}, \bibinfo {author}
  {\bibfnamefont {E.}~\bibnamefont {Zeldov}}, \bibinfo {author} {\bibfnamefont
  {D.~A.}\ \bibnamefont {Bonn}}, \bibinfo {author} {\bibfnamefont
  {R.}~\bibnamefont {Liang}}, \bibinfo {author} {\bibfnamefont {W.~N.}\
  \bibnamefont {Hardy}}, \ and\ \bibinfo {author} {\bibfnamefont {K.~A.}\
  \bibnamefont {Moler}},\ }\href
  {http://www.nature.com/nphys/journal/v5/n1/abs/nphys1127.html} {\bibfield
  {journal} {\bibinfo  {journal} {Nat. Phy.}\ }\textbf {\bibinfo {volume}
  {5}},\ \bibinfo {pages} {35} (\bibinfo {year} {2009})}\BibitemShut {NoStop}%
\bibitem [{\citenamefont {Yang}\ \emph {et~al.}(2012)\citenamefont {Yang},
  \citenamefont {Shen}, \citenamefont {Wang}, \citenamefont {Shan},
  \citenamefont {Ren},\ and\ \citenamefont {Wen}}]{Yang2012}%
  \BibitemOpen
  \bibfield  {author} {\bibinfo {author} {\bibfnamefont {H.}~\bibnamefont
  {Yang}}, \bibinfo {author} {\bibfnamefont {B.}~\bibnamefont {Shen}}, \bibinfo
  {author} {\bibfnamefont {Z.}~\bibnamefont {Wang}}, \bibinfo {author}
  {\bibfnamefont {L.}~\bibnamefont {Shan}}, \bibinfo {author} {\bibfnamefont
  {C.}~\bibnamefont {Ren}}, \ and\ \bibinfo {author} {\bibfnamefont {H.~H.}\
  \bibnamefont {Wen}},\ }\href {\doibase 10.1103/PhysRevB.85.014524} {\bibfield
   {journal} {\bibinfo  {journal} {Phys. Rev. B}\ }\textbf {\bibinfo {volume}
  {85}},\ \bibinfo {pages} {014524} (\bibinfo {year} {2012})}\BibitemShut
  {NoStop}%
\bibitem [{\citenamefont {Shapira}\ \emph {et~al.}(2015)\citenamefont
  {Shapira}, \citenamefont {Lamhot}, \citenamefont {Shpielberg}, \citenamefont
  {Kafri}, \citenamefont {Ramshaw}, \citenamefont {Bonn}, \citenamefont
  {Liang}, \citenamefont {Hardy},\ and\ \citenamefont
  {Auslaender}}]{Shapira2015}%
  \BibitemOpen
  \bibfield  {author} {\bibinfo {author} {\bibfnamefont {N.}~\bibnamefont
  {Shapira}}, \bibinfo {author} {\bibfnamefont {Y.}~\bibnamefont {Lamhot}},
  \bibinfo {author} {\bibfnamefont {O.}~\bibnamefont {Shpielberg}}, \bibinfo
  {author} {\bibfnamefont {Y.}~\bibnamefont {Kafri}}, \bibinfo {author}
  {\bibfnamefont {B.~J.}\ \bibnamefont {Ramshaw}}, \bibinfo {author}
  {\bibfnamefont {D.~A.}\ \bibnamefont {Bonn}}, \bibinfo {author}
  {\bibfnamefont {R.}~\bibnamefont {Liang}}, \bibinfo {author} {\bibfnamefont
  {W.~N.}\ \bibnamefont {Hardy}}, \ and\ \bibinfo {author} {\bibfnamefont
  {O.~M.}\ \bibnamefont {Auslaender}},\ }\href {\doibase
  10.1103/PhysRevB.92.100501} {\bibfield  {journal} {\bibinfo  {journal} {Phys.
  Rev. B}\ }\textbf {\bibinfo {volume} {92}},\ \bibinfo {pages} {100501}
  (\bibinfo {year} {2015})}\BibitemShut {NoStop}%
\bibitem [{\citenamefont {Albrecht}\ \emph {et~al.}(1991)\citenamefont
  {Albrecht}, \citenamefont {Gr{\"u}tter}, \citenamefont {Horne},\ and\
  \citenamefont {Rugar}}]{Albrecht1991}%
  \BibitemOpen
  \bibfield  {author} {\bibinfo {author} {\bibfnamefont {T.~R.}\ \bibnamefont
  {Albrecht}}, \bibinfo {author} {\bibfnamefont {P.}~\bibnamefont
  {Gr{\"u}tter}}, \bibinfo {author} {\bibfnamefont {D.}~\bibnamefont {Horne}},
  \ and\ \bibinfo {author} {\bibfnamefont {D.}~\bibnamefont {Rugar}},\ }\href
  {\doibase 10.1063/1.347347} {\bibfield  {journal} {\bibinfo  {journal} {J.
  Appl. Phys.}\ }\textbf {\bibinfo {volume} {69}},\ \bibinfo {pages} {668}
  (\bibinfo {year} {1991})}\BibitemShut {NoStop}%
\bibitem [{n:T()}]{n:Tip}%
  \BibitemOpen
  \href@noop {} {}\bibinfo {note} {We used five cantilevers: Numbers
  $1,~2,~4,~$ and $~5$ were by Nano-World, and number $~3$ by
  Paramount-Sensors. The spring constants were all $k=2-3$~N/m, and
  $f_0\approx79.5,~77.0,~93.7,~79.2,~71.3$~kHz for tips $\#1-\#5$
  respectively.}\BibitemShut {Stop}%
\bibitem [{\citenamefont {Tinkham}(1996)}]{Tinkham1996}%
  \BibitemOpen
  \bibfield  {author} {\bibinfo {author} {\bibfnamefont {M.}~\bibnamefont
  {Tinkham}},\ }\href@noop {} {\emph {\bibinfo {title} {Introduction to
  superconductivity}}}\ (\bibinfo  {publisher} {Courier Corporation},\ \bibinfo
  {year} {1996})\BibitemShut {NoStop}%
\bibitem [{\citenamefont {{Lan Luan}}\ \emph {et~al.}(2010)\citenamefont {{Lan
  Luan}}, \citenamefont {Auslaender}, \citenamefont {Lippman}, \citenamefont
  {Hicks}, \citenamefont {Kalisky}, \citenamefont {{Jiun Haw Chu}},
  \citenamefont {Analytis}, \citenamefont {Fisher}, \citenamefont {Kirtley},\
  and\ \citenamefont {Moler}}]{Luan2010}%
  \BibitemOpen
  \bibfield  {author} {\bibinfo {author} {\bibnamefont {{Lan Luan}}}, \bibinfo
  {author} {\bibfnamefont {O.~M.}\ \bibnamefont {Auslaender}}, \bibinfo
  {author} {\bibfnamefont {T.~M.}\ \bibnamefont {Lippman}}, \bibinfo {author}
  {\bibfnamefont {C.~W.}\ \bibnamefont {Hicks}}, \bibinfo {author}
  {\bibfnamefont {B.}~\bibnamefont {Kalisky}}, \bibinfo {author} {\bibnamefont
  {{Jiun Haw Chu}}}, \bibinfo {author} {\bibfnamefont {J.~G.}\ \bibnamefont
  {Analytis}}, \bibinfo {author} {\bibfnamefont {I.~R.}\ \bibnamefont
  {Fisher}}, \bibinfo {author} {\bibfnamefont {J.~R.}\ \bibnamefont {Kirtley}},
  \ and\ \bibinfo {author} {\bibfnamefont {K.~A.}\ \bibnamefont {Moler}},\
  }\href {http://link.aps.org/doi/10.1103/PhysRevB.81.100501} {\bibfield
  {journal} {\bibinfo  {journal} {Phys. Rev. B}\ }\textbf {\bibinfo {volume}
  {81}},\ \bibinfo {pages} {100501} (\bibinfo {year} {2010})}\BibitemShut
  {NoStop}%
\bibitem [{\citenamefont {Straver}\ \emph {et~al.}(2008)\citenamefont
  {Straver}, \citenamefont {Hoffman}, \citenamefont {Auslaender}, \citenamefont
  {Rugar},\ and\ \citenamefont {Moler}}]{Straver08}%
  \BibitemOpen
  \bibfield  {author} {\bibinfo {author} {\bibfnamefont {E.~W.~J.}\
  \bibnamefont {Straver}}, \bibinfo {author} {\bibfnamefont {J.~E.}\
  \bibnamefont {Hoffman}}, \bibinfo {author} {\bibfnamefont {O.~M.}\
  \bibnamefont {Auslaender}}, \bibinfo {author} {\bibfnamefont
  {D.}~\bibnamefont {Rugar}}, \ and\ \bibinfo {author} {\bibfnamefont {K.~A.}\
  \bibnamefont {Moler}},\ }\href {http://dx.doi.org/10.1063/1.3000963}
  {\bibfield  {journal} {\bibinfo  {journal} {Appl. Phys. Lett.}\ }\textbf
  {\bibinfo {volume} {93}},\ \bibinfo {pages} {172514} (\bibinfo {year}
  {2008})}\BibitemShut {NoStop}%
\bibitem [{\citenamefont {Zhang}\ \emph {et~al.}(2015)\citenamefont {Zhang},
  \citenamefont {Kim}, \citenamefont {Huefner}, \citenamefont {Ye},
  \citenamefont {Kim}, \citenamefont {Canfield}, \citenamefont {Prozorov},
  \citenamefont {Auslaender},\ and\ \citenamefont {Hoffman}}]{Zhang2015}%
  \BibitemOpen
  \bibfield  {author} {\bibinfo {author} {\bibfnamefont {J.~T.}\ \bibnamefont
  {Zhang}}, \bibinfo {author} {\bibfnamefont {J.}~\bibnamefont {Kim}}, \bibinfo
  {author} {\bibfnamefont {M.}~\bibnamefont {Huefner}}, \bibinfo {author}
  {\bibfnamefont {C.}~\bibnamefont {Ye}}, \bibinfo {author} {\bibfnamefont
  {S.}~\bibnamefont {Kim}}, \bibinfo {author} {\bibfnamefont {P.~C.}\
  \bibnamefont {Canfield}}, \bibinfo {author} {\bibfnamefont {R.}~\bibnamefont
  {Prozorov}}, \bibinfo {author} {\bibfnamefont {O.~M.}\ \bibnamefont
  {Auslaender}}, \ and\ \bibinfo {author} {\bibfnamefont {J.~E.}\ \bibnamefont
  {Hoffman}},\ }\href {\doibase 10.1103/PhysRevB.92.134509} {\bibfield
  {journal} {\bibinfo  {journal} {Phys. Rev. B}\ }\textbf {\bibinfo {volume}
  {92}},\ \bibinfo {pages} {134509} (\bibinfo {year} {2015})}\BibitemShut
  {NoStop}%
\bibitem [{\citenamefont {Park}\ \emph {et~al.}(2009)\citenamefont {Park},
  \citenamefont {Inosov}, \citenamefont {Niedermayer}, \citenamefont {Sun},
  \citenamefont {Haug}, \citenamefont {Christensen}, \citenamefont {Dinnebier},
  \citenamefont {Boris}, \citenamefont {Drew}, \citenamefont {Schulz},
  \citenamefont {Shapoval}, \citenamefont {Wolff}, \citenamefont {Neu},
  \citenamefont {Yang}, \citenamefont {Lin}, \citenamefont {Keimer},\ and\
  \citenamefont {Hinkov}}]{Park2009}%
  \BibitemOpen
  \bibfield  {author} {\bibinfo {author} {\bibfnamefont {J.~T.}\ \bibnamefont
  {Park}}, \bibinfo {author} {\bibfnamefont {D.~S.}\ \bibnamefont {Inosov}},
  \bibinfo {author} {\bibfnamefont {C.}~\bibnamefont {Niedermayer}}, \bibinfo
  {author} {\bibfnamefont {G.~L.}\ \bibnamefont {Sun}}, \bibinfo {author}
  {\bibfnamefont {D.}~\bibnamefont {Haug}}, \bibinfo {author} {\bibfnamefont
  {N.~B.}\ \bibnamefont {Christensen}}, \bibinfo {author} {\bibfnamefont
  {R.}~\bibnamefont {Dinnebier}}, \bibinfo {author} {\bibfnamefont {A.~V.}\
  \bibnamefont {Boris}}, \bibinfo {author} {\bibfnamefont {A.~J.}\ \bibnamefont
  {Drew}}, \bibinfo {author} {\bibfnamefont {L.}~\bibnamefont {Schulz}},
  \bibinfo {author} {\bibfnamefont {T.}~\bibnamefont {Shapoval}}, \bibinfo
  {author} {\bibfnamefont {U.}~\bibnamefont {Wolff}}, \bibinfo {author}
  {\bibfnamefont {V.}~\bibnamefont {Neu}}, \bibinfo {author} {\bibfnamefont
  {X.}~\bibnamefont {Yang}}, \bibinfo {author} {\bibfnamefont {C.~T.}\
  \bibnamefont {Lin}}, \bibinfo {author} {\bibfnamefont {B.}~\bibnamefont
  {Keimer}}, \ and\ \bibinfo {author} {\bibfnamefont {V.}~\bibnamefont
  {Hinkov}},\ }\href {\doibase 10.1103/PhysRevLett.102.117006} {\bibfield
  {journal} {\bibinfo  {journal} {Phys. Rev. Lett.}\ }\textbf {\bibinfo
  {volume} {102}},\ \bibinfo {pages} {117006} (\bibinfo {year}
  {2009})}\BibitemShut {NoStop}%
\bibitem [{\citenamefont {Ohgushi}\ and\ \citenamefont
  {Kiuchi}(2012)}]{Ohgushi2012}%
  \BibitemOpen
  \bibfield  {author} {\bibinfo {author} {\bibfnamefont {K.}~\bibnamefont
  {Ohgushi}}\ and\ \bibinfo {author} {\bibfnamefont {Y.}~\bibnamefont
  {Kiuchi}},\ }\href {\doibase 10.1103/PhysRevB.85.064522} {\bibfield
  {journal} {\bibinfo  {journal} {Phys. Rev. B}\ }\textbf {\bibinfo {volume}
  {85}},\ \bibinfo {pages} {064522} (\bibinfo {year} {2012})}\BibitemShut
  {NoStop}%
\bibitem [{\citenamefont {Kalisky}\ \emph {et~al.}(2011)\citenamefont
  {Kalisky}, \citenamefont {Kirtley}, \citenamefont {Analytis}, \citenamefont
  {Chu}, \citenamefont {Fisher},\ and\ \citenamefont {Moler}}]{Kalisky2011}%
  \BibitemOpen
  \bibfield  {author} {\bibinfo {author} {\bibfnamefont {B.}~\bibnamefont
  {Kalisky}}, \bibinfo {author} {\bibfnamefont {J.~R.}\ \bibnamefont
  {Kirtley}}, \bibinfo {author} {\bibfnamefont {J.~G.}\ \bibnamefont
  {Analytis}}, \bibinfo {author} {\bibfnamefont {J.-H.}\ \bibnamefont {Chu}},
  \bibinfo {author} {\bibfnamefont {I.~R.}\ \bibnamefont {Fisher}}, \ and\
  \bibinfo {author} {\bibfnamefont {K.~A.}\ \bibnamefont {Moler}},\ }\href
  {\doibase 10.1103/PhysRevB.83.064511} {\bibfield  {journal} {\bibinfo
  {journal} {Phys. Rev. B}\ }\textbf {\bibinfo {volume} {83}},\ \bibinfo
  {pages} {064511} (\bibinfo {year} {2011})}\BibitemShut {NoStop}%
\bibitem [{n:p()}]{n:pinning_forces}%
  \BibitemOpen
  \href@noop {} {}\bibinfo {note} {In the other samples, where we could not
  induce vortex motion, the forces we exerted while imaging were much higher
  and comparable to the forces reported in \PBa\ \cite{Yagil2016}.}\BibitemShut
  {Stop}%
\bibitem [{n:e()}]{n:estimatingF}%
  \BibitemOpen
  \href@noop {} {}\bibinfo {note} {In order to estimate
  $F_\mathrm{lateral}^\mathrm{max}$ we model the tip as an infinitely sharp
  truncated cone and integrate the known expression for the interaction energy
  of a magnet and a superconducting vortex \cite{Shapira2015}.}\BibitemShut
  {Stop}%
\bibitem [{\citenamefont {Shan}\ \emph {et~al.}(2012)\citenamefont {Shan},
  \citenamefont {Gong}, \citenamefont {Wang}, \citenamefont {Shen},
  \citenamefont {Hou}, \citenamefont {Ren}, \citenamefont {Li}, \citenamefont
  {Yang}, \citenamefont {Wen}, \citenamefont {Li},\ and\ \citenamefont
  {Dai}}]{Shan2012}%
  \BibitemOpen
  \bibfield  {author} {\bibinfo {author} {\bibfnamefont {L.}~\bibnamefont
  {Shan}}, \bibinfo {author} {\bibfnamefont {J.}~\bibnamefont {Gong}}, \bibinfo
  {author} {\bibfnamefont {Y.~L.}\ \bibnamefont {Wang}}, \bibinfo {author}
  {\bibfnamefont {B.}~\bibnamefont {Shen}}, \bibinfo {author} {\bibfnamefont
  {X.}~\bibnamefont {Hou}}, \bibinfo {author} {\bibfnamefont {C.}~\bibnamefont
  {Ren}}, \bibinfo {author} {\bibfnamefont {C.}~\bibnamefont {Li}}, \bibinfo
  {author} {\bibfnamefont {H.}~\bibnamefont {Yang}}, \bibinfo {author}
  {\bibfnamefont {H.~H.}\ \bibnamefont {Wen}}, \bibinfo {author} {\bibfnamefont
  {S.}~\bibnamefont {Li}}, \ and\ \bibinfo {author} {\bibfnamefont
  {P.}~\bibnamefont {Dai}},\ }\href {\doibase 10.1103/PhysRevLett.108.227002}
  {\bibfield  {journal} {\bibinfo  {journal} {Phys. Rev. Lett.}\ }\textbf
  {\bibinfo {volume} {108}},\ \bibinfo {pages} {227002} (\bibinfo {year}
  {2012})}\BibitemShut {NoStop}%
\bibitem [{\citenamefont {Hirsch}(1992)}]{Hirsch1992}%
  \BibitemOpen
  \bibfield  {author} {\bibinfo {author} {\bibfnamefont {J.}~\bibnamefont
  {Hirsch}},\ }\href
  {http://www.sciencedirect.com/science/article/pii/0921453492904159}
  {\bibfield  {journal} {\bibinfo  {journal} {Physica C}\ }\textbf {\bibinfo
  {volume} {199}},\ \bibinfo {pages} {305} (\bibinfo {year}
  {1992})}\BibitemShut {NoStop}%
\bibitem [{\citenamefont {Kogan}\ \emph {et~al.}(2009)\citenamefont {Kogan},
  \citenamefont {Martin},\ and\ \citenamefont {Prozorov}}]{Kogan2009}%
  \BibitemOpen
  \bibfield  {author} {\bibinfo {author} {\bibfnamefont {V.~G.}\ \bibnamefont
  {Kogan}}, \bibinfo {author} {\bibfnamefont {C.}~\bibnamefont {Martin}}, \
  and\ \bibinfo {author} {\bibfnamefont {R.}~\bibnamefont {Prozorov}},\ }\href
  {http://link.aps.org/doi/10.1103/PhysRevB.80.014507} {\bibfield  {journal}
  {\bibinfo  {journal} {Phys. Rev. B}\ }\textbf {\bibinfo {volume} {80}},\
  \bibinfo {pages} {014507} (\bibinfo {year} {2009})}\BibitemShut {NoStop}%
\bibitem [{\citenamefont {Chowdhury}\ \emph {et~al.}(2015)\citenamefont
  {Chowdhury}, \citenamefont {Orenstein}, \citenamefont {Sachdev},\ and\
  \citenamefont {Senthil}}]{Chowdhury2015}%
  \BibitemOpen
  \bibfield  {author} {\bibinfo {author} {\bibfnamefont {D.}~\bibnamefont
  {Chowdhury}}, \bibinfo {author} {\bibfnamefont {J.}~\bibnamefont
  {Orenstein}}, \bibinfo {author} {\bibfnamefont {S.}~\bibnamefont {Sachdev}},
  \ and\ \bibinfo {author} {\bibfnamefont {T.}~\bibnamefont {Senthil}},\ }\href
  {\doibase 10.1103/PhysRevB.92.081113} {\bibfield  {journal} {\bibinfo
  {journal} {Phys. Rev. B}\ }\textbf {\bibinfo {volume} {92}},\ \bibinfo
  {pages} {081113} (\bibinfo {year} {2015})}\BibitemShut {NoStop}%
\bibitem [{\citenamefont {Wang}\ \emph {et~al.}(2016)\citenamefont {Wang},
  \citenamefont {Abanov}, \citenamefont {Altshuler}, \citenamefont
  {Yuzbashyan},\ and\ \citenamefont {Chubukov}}]{Wang2016}%
  \BibitemOpen
  \bibfield  {author} {\bibinfo {author} {\bibfnamefont {Y.}~\bibnamefont
  {Wang}}, \bibinfo {author} {\bibfnamefont {A.}~\bibnamefont {Abanov}},
  \bibinfo {author} {\bibfnamefont {B.~L.}\ \bibnamefont {Altshuler}}, \bibinfo
  {author} {\bibfnamefont {E.~A.}\ \bibnamefont {Yuzbashyan}}, \ and\ \bibinfo
  {author} {\bibfnamefont {A.~V.}\ \bibnamefont {Chubukov}},\ }\href
  {https://link.aps.org/doi/10.1103/PhysRevLett.117.157001} {\bibfield
  {journal} {\bibinfo  {journal} {Phys. Rev. Lett.}\ }\textbf {\bibinfo
  {volume} {117}},\ \bibinfo {pages} {157001} (\bibinfo {year}
  {2016})}\BibitemShut {NoStop}%
\bibitem [{\citenamefont {Levchenko}\ \emph {et~al.}(2013)\citenamefont
  {Levchenko}, \citenamefont {Vavilov}, \citenamefont {Khodas},\ and\
  \citenamefont {Chubukov}}]{Levchenko2013}%
  \BibitemOpen
  \bibfield  {author} {\bibinfo {author} {\bibfnamefont {A.}~\bibnamefont
  {Levchenko}}, \bibinfo {author} {\bibfnamefont {M.~G.}\ \bibnamefont
  {Vavilov}}, \bibinfo {author} {\bibfnamefont {M.}~\bibnamefont {Khodas}}, \
  and\ \bibinfo {author} {\bibfnamefont {A.~V.}\ \bibnamefont {Chubukov}},\
  }\href {\doibase 10.1103/PhysRevLett.110.177003} {\bibfield  {journal}
  {\bibinfo  {journal} {Phys. Rev. Lett.}\ }\textbf {\bibinfo {volume} {110}},\
  \bibinfo {pages} {177003} (\bibinfo {year} {2013})}\BibitemShut {NoStop}%
\bibitem [{\citenamefont {Fernandes}\ \emph {et~al.}(2013)\citenamefont
  {Fernandes}, \citenamefont {Maiti}, \citenamefont {W\"olfle},\ and\
  \citenamefont {Chubukov}}]{Fernandes2013}%
  \BibitemOpen
  \bibfield  {author} {\bibinfo {author} {\bibfnamefont {R.~M.}\ \bibnamefont
  {Fernandes}}, \bibinfo {author} {\bibfnamefont {S.}~\bibnamefont {Maiti}},
  \bibinfo {author} {\bibfnamefont {P.}~\bibnamefont {W\"olfle}}, \ and\
  \bibinfo {author} {\bibfnamefont {A.~V.}\ \bibnamefont {Chubukov}},\ }\href
  {\doibase 10.1103/PhysRevLett.111.057001} {\bibfield  {journal} {\bibinfo
  {journal} {Phys. Rev. Lett.}\ }\textbf {\bibinfo {volume} {111}},\ \bibinfo
  {pages} {057001} (\bibinfo {year} {2013})}\BibitemShut {NoStop}%
\bibitem [{\citenamefont {Nomoto}\ and\ \citenamefont
  {Ikeda}(2013)}]{Nomoto2013}%
  \BibitemOpen
  \bibfield  {author} {\bibinfo {author} {\bibfnamefont {T.}~\bibnamefont
  {Nomoto}}\ and\ \bibinfo {author} {\bibfnamefont {H.}~\bibnamefont {Ikeda}},\
  }\href {\doibase 10.1103/PhysRevLett.111.167001} {\bibfield  {journal}
  {\bibinfo  {journal} {Phys. Rev. Lett.}\ }\textbf {\bibinfo {volume} {111}},\
  \bibinfo {pages} {167001} (\bibinfo {year} {2013})}\BibitemShut {NoStop}%
\bibitem [{\citenamefont {Shibauchi}\ \emph {et~al.}(2014)\citenamefont
  {Shibauchi}, \citenamefont {Carrington},\ and\ \citenamefont
  {Matsuda}}]{Shibauchi2014}%
  \BibitemOpen
  \bibfield  {author} {\bibinfo {author} {\bibfnamefont {T.}~\bibnamefont
  {Shibauchi}}, \bibinfo {author} {\bibfnamefont {A.}~\bibnamefont
  {Carrington}}, \ and\ \bibinfo {author} {\bibfnamefont {Y.}~\bibnamefont
  {Matsuda}},\ }\href
  {http://www.annualreviews.org/doi/abs/10.1146/annurev-conmatphys-031113-133921}
  {\bibfield  {journal} {\bibinfo  {journal} {Annu. Rev. Condens. Matter
  Phys.}\ }\textbf {\bibinfo {volume} {5}},\ \bibinfo {pages} {113} (\bibinfo
  {year} {2014})}\BibitemShut {NoStop}%
\bibitem [{\citenamefont {Chowdhury}\ \emph {et~al.}(2013)\citenamefont
  {Chowdhury}, \citenamefont {Swingle}, \citenamefont {Berg},\ and\
  \citenamefont {Sachdev}}]{Chowdhury2013}%
  \BibitemOpen
  \bibfield  {author} {\bibinfo {author} {\bibfnamefont {D.}~\bibnamefont
  {Chowdhury}}, \bibinfo {author} {\bibfnamefont {B.}~\bibnamefont {Swingle}},
  \bibinfo {author} {\bibfnamefont {E.}~\bibnamefont {Berg}}, \ and\ \bibinfo
  {author} {\bibfnamefont {S.}~\bibnamefont {Sachdev}},\ }\href {\doibase
  10.1103/PhysRevLett.111.157004} {\bibfield  {journal} {\bibinfo  {journal}
  {Phys. Rev. Lett.}\ }\textbf {\bibinfo {volume} {111}},\ \bibinfo {pages}
  {157004} (\bibinfo {year} {2013})}\BibitemShut {NoStop}%
\bibitem [{\citenamefont {Gordon}\ \emph
  {et~al.}(2010{\natexlab{b}})\citenamefont {Gordon}, \citenamefont {Kim},
  \citenamefont {Tanatar}, \citenamefont {Prozorov},\ and\ \citenamefont
  {Kogan}}]{Gordon2010b}%
  \BibitemOpen
  \bibfield  {author} {\bibinfo {author} {\bibfnamefont {R.~T.}\ \bibnamefont
  {Gordon}}, \bibinfo {author} {\bibfnamefont {H.}~\bibnamefont {Kim}},
  \bibinfo {author} {\bibfnamefont {M.~A.}\ \bibnamefont {Tanatar}}, \bibinfo
  {author} {\bibfnamefont {R.}~\bibnamefont {Prozorov}}, \ and\ \bibinfo
  {author} {\bibfnamefont {V.~G.}\ \bibnamefont {Kogan}},\ }\href {\doibase
  10.1103/PhysRevB.81.180501} {\bibfield  {journal} {\bibinfo  {journal} {Phys.
  Rev. B}\ }\textbf {\bibinfo {volume} {81}},\ \bibinfo {pages} {180501}
  (\bibinfo {year} {2010}{\natexlab{b}})}\BibitemShut {NoStop}%
\end{thebibliography}%

\end{document}